# Theoretical Ion Sputtering Yields from Loose Powders using a Multiscale Monte Carlo Approach

S. Verkercke[1,2,3*], D. Berhanu[4], C. Bu[5], B. Clouter-Gergen[6], F. Leblanc[7], J. R. Lewis[6], L. S. Morrissey[6,8*], and D. W. Savin[5*]

[1] LATMOS/CNRS, Université Versailles Saint Quentin, Guyancourt, France.

[2] Centre National d'Etudes Spatiales, Paris, France

[3] Laboratoire de Physique des Plasmas (LPP), CNRS, Ecole polytechnique, Institut Polytechnique de Paris, 91120 Palaiseau, France

[4] Department of Science and Mathematics, Fashion Institute of Technology, New York, NY 10001, USA.

[5] Columbia Astrophysics Laboratory, Columbia University, New York, NY 10027, USA.

[6] Memorial University of Newfoundland, St. John's NLA1C 5S7, Canada.

[7] LATMOS/CNRS, Sorbonne Université, Paris, France.

[8] American Museum of Natural History, New York, NY 10024, USA.

*Corresponding Authors: sebastien.verkercke@latmos.ipsl.fr; lsm088@mun.ca; dws26@columbia.edu.

Abstract

Ion sputtering from loose powders remains poorly understood despite its relevance to planetary science and industry. We developed a multiscale Monte Carlo model to simulate sputtering from powders, using a higher-fidelity approach for the target geometry compared to voxel-based methods. Simulating $Kr^+$ ions impacting Cu powders and flat slabs, we show that sputtering from loose powders differs markedly from that of flat slabs or rough surfaces. The main differences are: (1) for incident angles $\alpha > 0°$ relative to the bulk normal, the escaping sputtering yield is dominated by backward-directed ejecta for all ion energies; (2) for $\alpha \leq 60°$, the yield peaks toward the ion-beam origin, similar to the opposition effect seen in optical observations of airless bodies; (3) the angular distribution peak is half or less than that of a flat slab; (4) as ion energy increases, no evolution occurs from primary to secondary knock-on sputtering in the ejecta angular distribution. We attribute these behaviors to the powder's interconnected voids. Ions penetrate these voids and sputter underlying grains; the ejecta then preferentially escape toward the ion-beam origin, where shadowing is minimal. We derive two fitting functions: 1) relating the escaping sputtering yield of a powder to that of a flat surface, depending only on porosity, incident angle, mean local incidence angle, and the corresponding flat slab yield; 2) providing the double-differential angular distribution of the escaping ejecta for porosities $\geq 0.49$. These provide a potentially universal fitting function of the absolute doubly-differential escaping sputtering yield from loose powders.

1. Introduction

Energetic ions impacting a surface can lead to the ejection of atoms in a process known as ion sputtering (Lehmann & Sigmund 1966, Sigmund 1969, 2005). While this process has been extensively studied on flat surfaces (Behrisch 1981, Behrisch & Eckstein 2007, Rauschenbach 2022), the current understanding of sputtering from powders and porous matter remains limited, due to the paucity of theoretical and experimental results (as we briefly review below). However, sputtering from these surfaces is relevant to a wide range of fields including solar wind ion interactions with the regolith surfaces of airless planetary bodies such as the Moon and Mercury (Leblanc et al. 2022, Teolis et al. 2023), semiconductor manufacturing (Depla 2021, Ohshima 2023, Yu et al. 2024) and fusion reactor wall damage forming a "fuzzy" surface (Abernethy 2017, Wright 2022). Here, we focus our attention on sputtering from loose powders, though our work and results can readily be extended to porous solids, i.e. fuzzy surfaces, as well as to non-porous rough surfaces.

The earliest theoretical models that we are aware of for sputtering from loose powders are the simple analytical models of Hapke (1986) and Johnson (1989), which were roughly reproduced by the subsequent Monte Carlo simulations of Cassidy & Johnson (2005).  This latter work more accurately accounted for the granular nature of the surface (hereafter, granular roughness) and the porosity (i.e., the volume of the powder not occupied by grains), though they did not consider any surface roughness of the individual grains.  More importantly, in order to make the calculations readily tractable, that study made a number of simplifying assumptions for the angular distribution of the ejecta.  Their model predicted an ejecta retention rate of ~ 72 % for a cosine-law-dependence for the polar angle of the ejecta distribution and increased to ~ 92% for a forward-directed distribution, assuming a sticking coefficient of the ejecta onto the grains of $S = 1$ for both cases.  At the time of that work, there were no quantitative experimental results for comparison.  The only available experimental results were the qualitative results of Hapke & Cassidy (1978), who conducted experiments with basalt-like glass and observed that more than 90% of the sputtered material remained within the granular target.  So while there does appear to be rough qualitative agreement between the theoretical and experimental results, it is difficult to reliably apply the theoretical results of any of the above three models to real world situations.

It is only recently that there have been further theoretical studies of ion sputtering from loose powders (Brötzner et al. 2025 and our work here). These have grown out of related theoretical studies into energetic neutral atom (ENA) emission from the lunar regolith due the neutralization and reflection of solar wind protons (Szabo et al. 2022a, 2023; Leblanc et al. 2023; Verkercke et al. 2023). Two complementary approaches have been implemented. Each approach splits the problem into three parts: generating the regolith structure, calculating the ejecta yield and distribution, and tracing the trajectories of the ejecta within and out of the structure.

In order to generate the regolith structure for the simulations, both groups dropped spherical grains into a sample volume. However, they treated the interaction between the grains differently.  Szabo et al. (2022a, 2023) assumed a sticking coefficient of 1 to build up their structure. This is computationally efficient, but does not necessarily generate a realistic regolith structure. This contrasts with Leblanc et al. (2023) and Verkercke et al. (2023) who used a molecular dynamics (MD) grain-to-grain contact model. Though this takes longer computationally, it uses physical principals and is expected to produce a more realistic regolith structure.

The backscattering (i.e., reflection) coefficients and the angular distributions were calculated by Szabo et al. (2022a, 2023) using SDTrimSP-3D (von Toussaint et al. 2017; which is built upon SDTrimSP, Mutzke et al. 2014), and by Leblanc et al. (2023) and Verkercke et al. (2023) using a combination of SDTrimSP and MD. SDTrimSP uses the binary collision approximation (BCA). MD takes many-body interactions into account. Good agreement has been found for the SDTrimSP and MD approaches for sputtering yields and reflection coefficients (Verkercke et al. 2023), but to our knowledge, no study has ever compared both approaches for the angular distributions of the sputtering yield or of the reflected atoms.

An important difference between the approaches of these two groups is the way that the surfaces are handled. SDTrimSP-3D discretizes the structure into cubic cells (dubbed voxels). Within this orthogonal approximation to smooth surfaces, von Toussaint et al. has shown that the total sputtering yield can be well reproduced using voxels less than ~ 10 nm on each side. However, they did not report on how well the angular distribution of the ejecta was reproduced. This is an important issue for accurately predicting the trajectories of the ejecta as it travels within and out of the powder structure. Another limitation of SDTrimSP-3D is that the volume that is computationally feasible to simulate is constrained by the voxel size. For example, Szabo et al. (2022a) used grain sizes of 0.3 µm and voxels of 4 nm, creating sample cells of 1.2 µm by 1.2 µm by 0.9 – 2.7 µm containing 50 – 100 grains with periodic boundaries; but the size of the sample and of the grains are not representative of the typical grain sizes of ~ 50-100 µm measured in the Apollo samples (McKay et al. 1991). So it is unclear how applicable the results of SDTrimSP-3D are to planetary regolith simulations.

This contrasts with the approach of Leblanc et al. (2023) and Verkercke et al. (2023) who used a combination of SDTrimSP and MD. They calculated the local angle of incidence for each impact using the exact surface structure. The ejecta angular distribution predicted by SDTrimSP has been recently benchmarked by the absolute doubly differential angular sputtering yields measurements of Bu et al. (2024) and found to be in good agreement. Furthermore, because the simulation volumes of Leblanc et al. (2023) and Verkercke et al. (2023) were not constrained by the sputtering model, they were able to simulate a more realistic volume of 1 mm × 1 mm × 1 – 12 mm containing 12,000 to 800,000 grains with a more planetary-science-relevant grain size distribution ranging from 6 to 700 µm. As is typical, periodic boundaries were used.

Ray-tracing was used to follow the trajectories of the ejecta. In both approaches, a direction vector was assigned for each ejected atom and all trajectories were considered as linear. If an atom encountered a grain in the powder, it was considered as redeposited. Those atoms that escaped were mapped according to their polar and azimuthal angles. We note that in addition to SDTrimSP-3D (von Toussaint et al. 2017) and our work here, other groups have also presented ray-tracing methods for ejected atoms (e.g., Cassidy & Johnson 2005, Möller 2014, Li et al. 2016, Cupak et al. 2021).

The previous theoretical studies relying on SDTrimSP-3D to model ENA reflected by the porous lunar regolith predicted a total ENA fraction (Szabo et al. 2022a) in good agreement with Chandrayaan-1 spacecraft measurements (Futaana et al. 2012). However, those studies did not accurately reproduce the ENA backscattering angular distributions (Szabo et al. 2023) measured by Chandrayaan-1 (Schaufelberger et al. 2011). This might be due to the maximum sample size SDTrimSP-3D is capable of simulating or to how well SDTrimSP-3D reproduces the angular distribution of the ejecta. Theoretical models combining an MD-simulated regolith with a treatment of the ion impacts on grains using SDTrimSP/MD have been found to provide results similar to SDTrimSP-3D for the total ENA reflected fraction, while better reproducing the angular distribution observed at the Moon (Verkercke et al. 2023).

Here we present an ion sputtering model for a loose powder that builds on the approach of Leblanc et al. (2023) and Verkercke et al. (2023). Our approach is applicable in general to loose powders, porous solids, and non-porous rough surfaces, though our focus here is on loose powders. For examples of recent theoretical work concerning sputtering from non-porous rough surfaces, we point the reader to von Toussaint et al. (2017), Cupak et al. (2021), Szabo et al. (2022b), and Biber et al. (2022). More specifically, we study the polar and azimuthal angular distribution of the absolute sputtering yield produced by a $Kr^+$ beam impacting a Cu powder. We investigate the dependencies of the yield on incidence angle and energy of the Kr beam. This choice of ion and target provides a direct comparison with recent laboratory experimental results for both a flat slab (Bu et al. 2024) and a loose powder (Bu et al. in preparation). This comparison will enable us to study the retention effect of a granular surface and to develop a benchmarked theoretical scaling relating the sputtering yield of a granular surface to the sputtering yield from a flat surface.

The rest of this paper is organized as follows: The simulation set-up is detailed in Section 2. Section 3 explores the influence of the different ion beam parameters such as the incidence energy and angle, as well as the influence of the structure of the powder. Section 4 presents a discussion on our simulation results and Section 5 presents our conclusions.

2. Simulation setup

We have developed a Loose Powder Sputtering Simulation (LooPSS) to calculate the sputtering yield and angular distribution due to ions impacting a powder. The simulation divides the problem up into three parts: using MD to generate the powder target; using a BCA code to calculate the needed sputtering yields and angular distributions; and following the trajectories of the ejecta within and out of the powder structure. The first and third steps operate on the granular scale, while the second step operates on the atomic scale. Monte Carlo simulations are used for each steps of the simulation, as we explain below.

2.1 Granular packing

The granular packing is calculated using the GRANULAR module of the open-source code Large-scale Atomic/Molecular Massively Parallel Simulator (LAMMPS; Thompson et al. 2022). The LAMMPS simulation box was established with a 1 mm × 1 mm square bottom along the *xy* plane at *z* = 0, with the box height along the +*z*- axis. Boundary conditions are periodic in *x* and *y* and fixed in *z*, simulating an infinite slab of powder. A total of 8000 spherical grains, with a uniform grain size distribution between 50 and 90 µm, were dropped randomly in *x*, *y*, and *z* from a region between 3 mm < *z* < 4.5 mm above the bottom of the simulation box. The grains had no initial velocity and were subject only to an Earth-like gravity. All grains reaching the bottom of the simulation box (*z* = 0) were considered as fixed for the rest of the simulation while the rest of the grains were free to move until finding an equilibrium position. We use a Johnson–Kendall–Roberts (JKR) contact model to account for a realistic description of grain-on-grain contacts (Johnson et al. 1971, Verkercke et al. 2023). The JKR model builds upon a classic Hertz contact model that includes the effects of adhesive contact forces between grains in contact (Johnson et al. 1971). The simple Hertz contact model describes the normal and shear forces based on theoretical analysis of the deformation of smooth, elastic spheres in frictional contact (Mindlin & Deresiewicz 1953; Elata & Berryman 1996). The JKR model extends this description by incorporating an additional attractive force ($F_a$) between grains, as such a force was observed experimentally (Johnson et al. 1971). For spherical grains, this additional force acts on the circular contact area characterized by a radius, *a*, between the two elastic solid grains. This radius depends on the radii and elastic constants of the grains, and on the load acting on the grains. At equilibrium, the adhesion force can be expressed as:

$$F_a = 2 \pi \gamma a \qquad (1)$$

where $\gamma$ is the surface energy per unit of contact area. For our case of spherical, pure Cu grains, we use $\gamma = 1.8$ J/m$^2$ (Sheng et al. 2011).

When all the grains reached an equilibrium, the powder had a height of 4.2 mm and a bulk density of 2.87 g/cm$^3$, corresponding to a porosity $p$ = 0.68. No sintering was modeled as the sample was considered at room temperature ($T$ ~ 298 K; Jang & Ahn 2023). This structure was then "shaken" to simulate the handling of the sample in the laboratory. We applied random forces simultaneously on each grain in the *xy* plane for 2 µs followed by a rest period of 0.1 ms. The forces are randomly selected from a uniform distribution between 0 and 10$^{-7}$ N, which corresponds to the typical acceleration of human arm movements (Gaveau & Papaxanthis 2011). This sequence was repeated 100 times. The height of the sample after the shaking sequence was 3.3 mm. The top 600 µm layer formed a very loose stacking of grains at the surface, which is commonly called a fairy-castle-like structure (Szabo et al. 2022a, Verkercke et al. 2023). We removed this layer to simulate the scraping off of the real sample used in the experiment. We also removed the bottom 200 µm layer, which was affected by the fixed bottom condition. The final height of the sample was 2.5 mm with a bulk density of 4.48 g/cm$^3$ and porosity of 0.49 (Fig. 1). These quantities of the simulated Cu powder target are very close to the density of 4.78 ± 0.24 g/cm$^3$ and porosity of 0.46 ± 0.03 as measured by Bu et al. (in preparation).

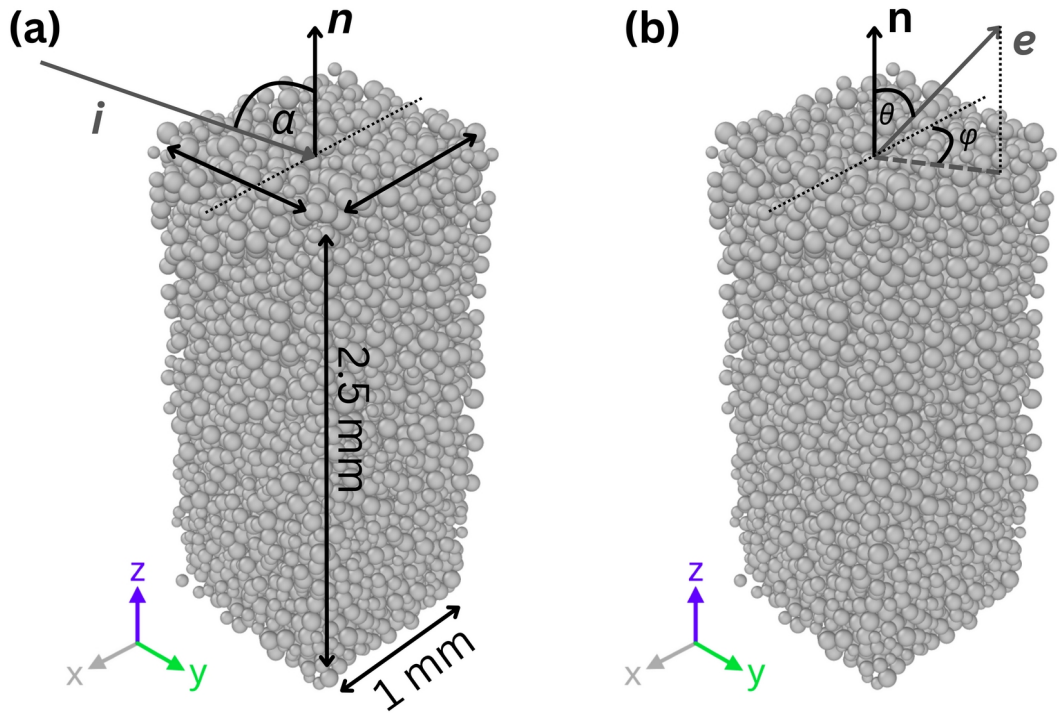

**Fig. 1.** Cu powder structure created using the LAMMPS GRANULAR package. The simulated volume is 1 mm × 1 mm × 2.5 mm and the powder has a porosity of 0.49. (a) The angle $\alpha$ lies between the bulk surface normal $\mathbf{n}$ and the incidence ion vector $\mathbf{i}$. The black dotted line is the projection of the $\mathbf{i}$ vector onto the $xy$ plane. (b) The polar angle $\theta$ is defined as the angle between $\mathbf{n}$ and the yield directional vector $\mathbf{e}$. The gray dashed line is the projection of $\mathbf{e}$ onto the $xy$ plane. Lastly, the azimuthal angle $\varphi$ is defined as the angle between the respective $\mathbf{i}$ and $\mathbf{e}$ vectors projection on the $xy$ plane, as measured from the $-x$ axis.

2.2 Ion on grain interaction

Simulations of the ion-grain interactions were conducted using the Monte Carlo model SDTrimSP (Version 7.00; Mutzke et al. 2024), which is an extension of the Transport of Ions in Matter (TRIM) software. SDTrimSP can be run in static (S) or dynamic (D) modes using serial (S) or parallel (P) processing. SDTrimSP is a BCA model that considers sputtering to be the result of a cascade of binary collisions, eventually leading to the ejection of an atom from the surface. It tracks the kinetic energy of the sputtered and substrate atoms but does not consider the effects of charge exchange, potential sputtering, substrate crystallinity, or surface roughness. SDTrimSP has been used extensively to study sputtering for experimental and planetary cases (Höfsass et al. 2014, Szabo et al. 2022a, Morrissey et al. 2023, Jäggi et al. 2024). In our previous study of $Kr^+$ impacting a flat Cu slab, the total sputtering yield and doubly differential angular distributions of the yield from SDTrimSP simulations agreed well with the experimental results (Bu et al. 2024).

SDTrimSP simulations were conducted for Kr impacting a flat Cu surface. We used incidence energies, $E_i$, of 1, 5, and 20 keV and incidence angles, $\theta_i$, between 0° and 90° in 5° increments. For each $E_i$ and $\theta_i$ pair, a total of 100 000 Kr impacts were simulated onto a 1000 Å thick flat Cu slab. As described in Morrissey et al (2022) and Bu et al. (2024), static simulations were conducted, using the Cu mono-atomic cohesive energy for the surface binding energy of Cu, and with serial processing. We tracked the ejection energy and polar and azimuthal emission angles for each sputtered Cu atom. We binned the ejecta in direction in 5° × 5° bins in both polar and azimuthal angles and used the resulting data set to compute the polar and azimuthal angular distributions of the ejecta as a function $E_i$ and $\theta_i$. Fig. 2 presents the ejecta angular distributions for representative sets of $E_i$ and $\theta_i$. The cumulative distribution functions (CDF) of the polar and azimuthal angles can then be computed from these data for each pair of $E_i$ and $\theta_i$, which we later use in a Monte Carlo approach to determine the yield as described later in Section 2.3.

We note that the results shown in Fig. 2 demonstrate the well-known sputtering behavior of flat slabs (Behrisch & Eckstein 2007). At a low energy (here 1 keV) with increasing $\alpha$, the ejecta from a flat surface is increasingly forward directed (relative to the incident ion). The sputtering process is dominated by primary knock-on collisions (defined as those collisions that still have memory of the momentum of the incident ion). Atoms are ejected through a direct collision with the incident ion, which occurs favorably as the penetration depth of the ion shallow. At a higher energy (here 5 keV), the penetration depth increases and secondary knock-on collisions come to dominate the sputtering process, bringing the ejecta distribution closer to cylindrically symmetric. The incident ion scatters multiple times in the substrate, randomizing the process and producing a cascade of collisions that erase any memory of the incident momentum of the ions. But as the impact angle increases, the ejecta still becomes forward directed as the penetration depth decreases, thereby favoring primary knock-on collisions. Increasing to a yet-higher energy (here 20 keV) brings the ejecta even closer to cylindrical symmetry even at large $\alpha$ as the penetration depth and the number of secondary knock-on collisions increases. At all energies, the results exhibit a mirror symmetry in the incidence plane, defined as the plane containing the ion beam and the surface normal. BCA model results such as these for flat slabs have been extensively benchmarked experimentally for the total sputtering yield and for the angular distribution of the ejecta in the incidence plane, i.e., as a function of the polar angle of the ejecta (Behrisch & Eckstein 2007). We are aware of only a few such benchmark studies for the doubly differential angular sputtering yield, i.e., as a function of both the polar and azimuthal angles of the ejecta (Becerra-Acevedo et al. 1984, Bu et al. 2024).

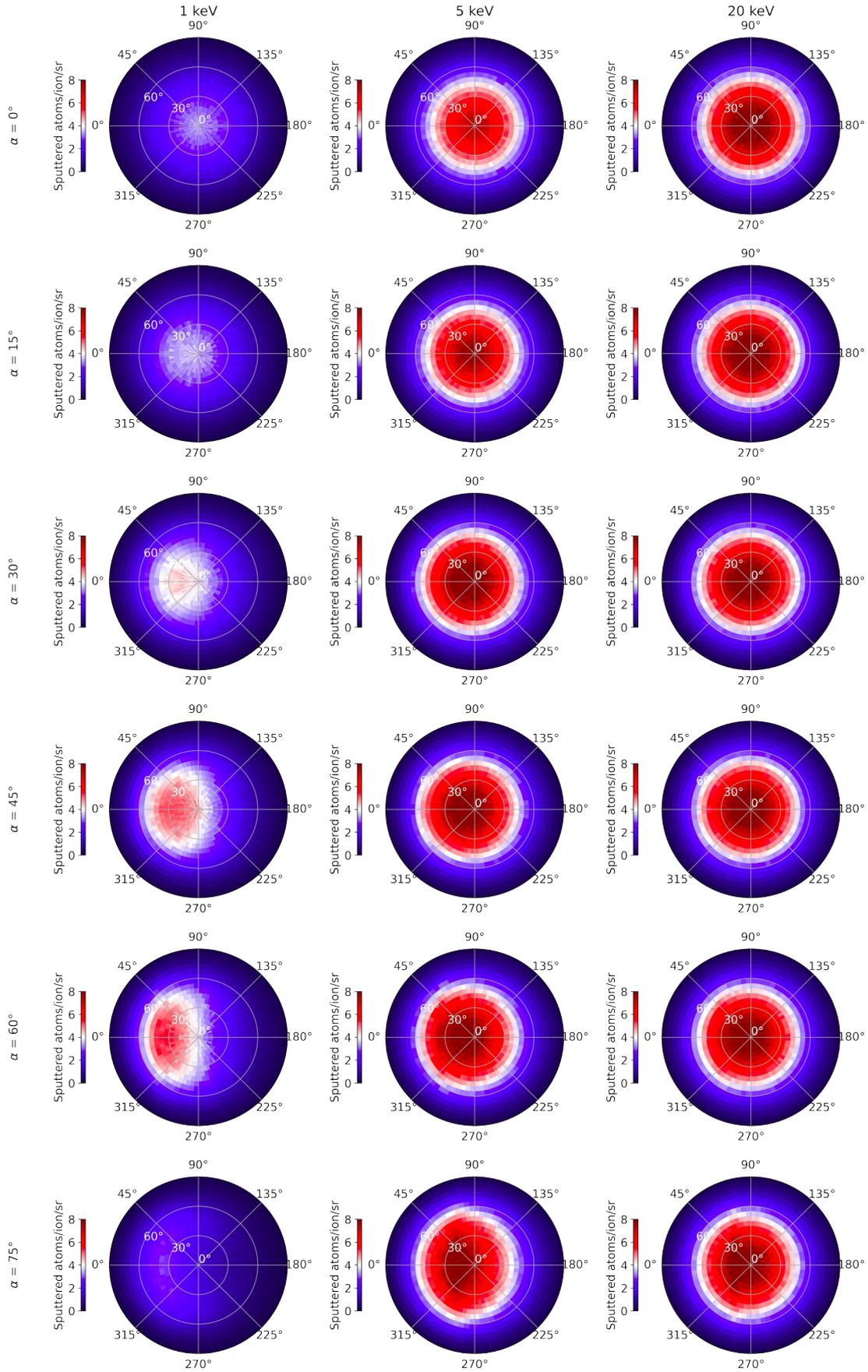

**Fig 2**. Sputtering yield angular distribution for a Kr$^+$ ion beam on a Cu slab as a function of the beam energy and of the incidence angle. The ion beam is directed along $\varphi = 0°$ with an incident polar angle of $\alpha$.

2.3 Tracking the ions and the yield

The incident Kr ions were launched downward from a height equal to the highest grain plus the radius of the largest grain, ensuring that no ion is initially created inside a grain. The time step of the simulation was chosen so that the spatial step did not exceed 1 µm, which is small compared to the grain size distribution. A global incidence angle $\alpha$ with respect to the bulk normal was applied to the Kr ions as shown in Fig. 1(a). The incidence plane is defined as the plane containing the ion beam and the bulk surface normal, which is the $xz$ plane in Fig. 1(a). Ions launched at $\alpha \neq 0°$ have a non-zero velocity component in the $-x$ direction.

Prior to launching the ions, a 3D grid was applied to the powder model. The grid was constructed of cubic cells, with sides equal to the diameter of the largest grain. This ensured that a grain can only overlap with a maximum of 8 cells. As an ion evolved towards the powder, the model evaluated its current location in the grid and identified the cell containing the ion. An impact occurred if the distance between the ion and one of these grains in this cell was smaller than that grain's radius.

Fig. 3(a) shows the local geometry of an impact on a grain. The local normal vector, shown as $z'$ in the panel, is defined as the normalized vector formed by the grain center and the impact point. The local incidence angle, $\alpha'$, is the angle between $z'$ and the Kr incidence vector, $i$. Since the scale of each impact is much smaller than the size of a grain, we approximated each impact as occurring on a flat surface, enabling us to use the SDTrimSP data described in Section 2.2. Knowing the local incidence angle, a linear interpolation was performed between the SDTrimSP yield values corresponding to the two closest angles in the data set given by $\theta_{i,k} < \alpha' < \theta_{i,k+1}$, where $k$ is the index of the CDF, as shown in Fig. 3(b). This enables us to determine the corresponding angular sputtering yields for any local incidence angle.

For each impact, the yield computed through the linear interpolation is rounded to the closest integer, representing the number of Cu atoms sputtered by the Kr ion. If the yield is much smaller than one, it is multiplied by an appropriate factor to make it greater than 1. For those cases, the total number of ions is also multiplied by the same factor after the simulation to account for the low efficiency of the sputtering. For each sputtered atom, the ejecta polar and azimuthal angles were determined through a Monte Carlo sampling of a CDF derived from a linear interpolation of the SDTrimSP CDFs (cf., Section 2.2). We randomly selected a number between 0 and 1 for each angle, and determined the corresponding value for each angle from the CDF. Each sputtered Cu atom was defined by these angles. As shown in Fig. 3(c), the local polar angle $\theta'$ was defined with respect to the local normal to the grain surface; the local azimuth angle $\varphi'$ was defined with respect to the projection of $i$ onto the local plane of impact on the grain. Using $\theta'$ and $\varphi'$, the direction of the ejected Cu atom was determined in the local reference frame $O'(x'y'z')$. Fig. 3(d) shows the ejection vector $e$ in the local reference frame. By transforming the vector from the local reference frame to the bulk reference frame, one can then follow the trajectory of each atom as it moves out of the powder, as shown in Fig. 1(b). If the sputtered Cu atom reached the plane from where the Kr ions were launched, the Cu atom was considered to have escaped out the powder as shown by the $e$ vector in Fig. 3(d). However, if the sputtered atom interacted with another grain, it was considered to be shadowed by the powder as shown by the $e'$ in Fig. 3(d). The ejecta trajectories were determined with a ray-tracing method similar to that described for the incident Kr ions. The azimuthal distribution of the ejected atoms can be divided in two half-hemispheres along the $y$-axis as shown in Fig. 1(b) – the backward (90° < $\varphi$ < 270°) and forward (-90° ≤ $\varphi$ < 270°) half-hemispheres.

We note that an ion can also be reflected after a collision with a grain. The probability of it being reflected and the reflection energy after the impact were computed using SDTrimSP. Reflected ions can subsequently cause sputtering if the ions are energetic enough. Reflected ions were tracked similarly to the primary ions and their sputtering contribution included in the reported results, but these contributed to less than 2% of the total sputtering yield.

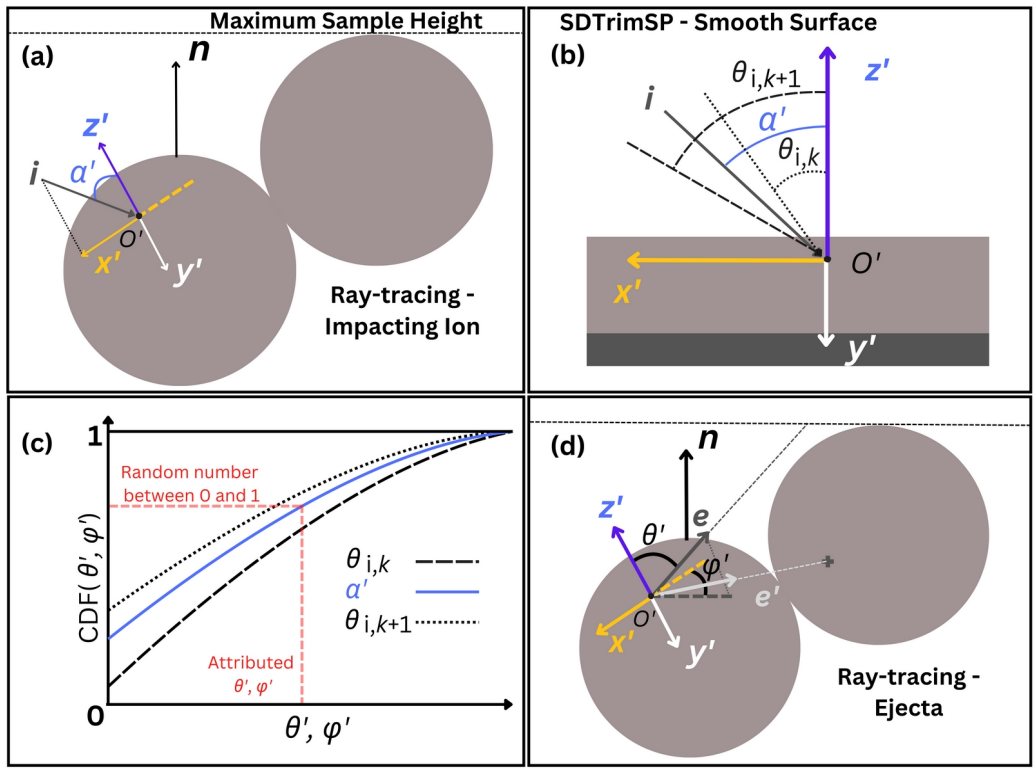

**Fig. 3.** Diagram representing the sputtering model. (a) Ray-tracing and local geometry of an ion impact on a grain. The vector ***n*** is the normal to the sample bulk, ***O'*** is the impact point, ***z'*** is the local normal on the grain, ***x'*** is the opposite of the projection of the incidence vector ***i*** in the impact plane (the plane defined by ***i*** and ***z'***), ***y'*** is a vector perpendicular to ***x'*** and ***z'***, and $α'$ is the local incidence angle between ***i*** and ***z'***. (b) Local incidence angle $α'$, approximated as occurring on a flat surface, and the two bracketing angles $θ_{i,k}$ and $θ_{i,k+1}$ used for SDTrimSP calculations. (c) Representation of the CDFs for $θ'$ and $φ'$ computed using SDTrimSP on a smooth surface. The appropriate $α'$ CDF is computed using a linear interpolation between the CDFs for $θ_{i,k}$ and $θ_{i,k+1}$. A Monte-Carlo sampling of this CDF gives the ejection vector, ***e***, for each sputtered atom. (d) Geometry of the ray-tracing of sputtered atoms, with an example of the two possible outcomes: the ***e*** vector shows the trajectory of an escaping atom and the ***e'*** vector shows the trajectory of a shadowed (i.e., retained) atom.

3. Results

We have investigated the effect of the incidence energy and of the incidence angle on the doubly-differential angular yield of escaped atoms (defined as escaped-atoms/ion/sr; hereafter, interchangeably called doubly-differential angular escaping sputtering yield), the retention yield (defined as retained-atoms/ion), the total sputtering yield (which is the sum of the escaping and retention yields and defined as sputtered-atoms/ion), and the escape percentage (defined as escaped-atoms/sputtered-atoms). The Kr ions energies considered were 1, 5, and 20 keV. Incidence angle were considered of $\alpha$ = 0°, 15°, 30°, 45°, 60°, and 75°. For each energy, 100 000 ions were launched towards the Cu powder and the resulting ejecta angular distributions were computed.

The angular distribution of the yield escaping from the Cu powder is shown in Fig. 4 for each incident angle and energy pair, showing mirror symmetry in the incidence play. Significant differences relative to a flat slab can be readily seen by comparing to the results in Fig. 2. These difference are highlighted in Fig. 5, which plots the doubly differential angular escaping yields in the incidence plane for the slab and powder results shown in Figs. 2 and 4, respectively. The most pronounced differences are that (1) the escaped sputtering yield from a powder for $\alpha > 0°$ is predominantly in the backward direction (relative to the incident ion) for all impact energies; (2) the yield is peaked toward the ion beam origin for $\alpha \leq 60°$, suggesting a mechanism similar to the opposition effect or opposition surge that is seen in observations of airless planetary bodies when the angle formed by the illumination source, the object, and the observer (i.e., phase angle) is $\lesssim 5°$ (e.g., Gehrels 1956, 1957, 1964, Hapke 1963); (3) the peak in the angular distribution is typically half or less than that of a flat slab; and (4) there is no evidence in the angular distribution of any evolution from primary knock-on sputtering to secondary knock-on sputtering with increasing ion energy. Similar results were found in the powder measurements of Bu et al. (in preparation) for (1), (2), and (3). However, those measurements were only for 20 keV Kr$^+$ and did not explore the incident energy dependence of the ejecta distribution.

Table 1 reports various integrals of our model results. The total escaping yield is given by $Y_E$ and the total retention yield by $Y_R$. The escape percentage out of the powder is given by $100 Y_E/(Y_E + Y_R)$. The directionality of the yield is given by the ratio $R_E$ of the backward-to-forward escaping atoms and is defined by integrating the doubly differential escaping sputtering yields over $\theta$ and respectively from 270° $\leq \varphi <$ 90° and 90° $\leq \varphi <$ 270°, as defined by Schaufelberger et al. (2011). The total sputtering yield from a flat sample is given by $Y_F$ and its directionality is given by the ratio $R_F$. Lastly, the yield ratio $Y_E/Y_F$ characterizes the difference between the ejecta from a powder and from a slab.

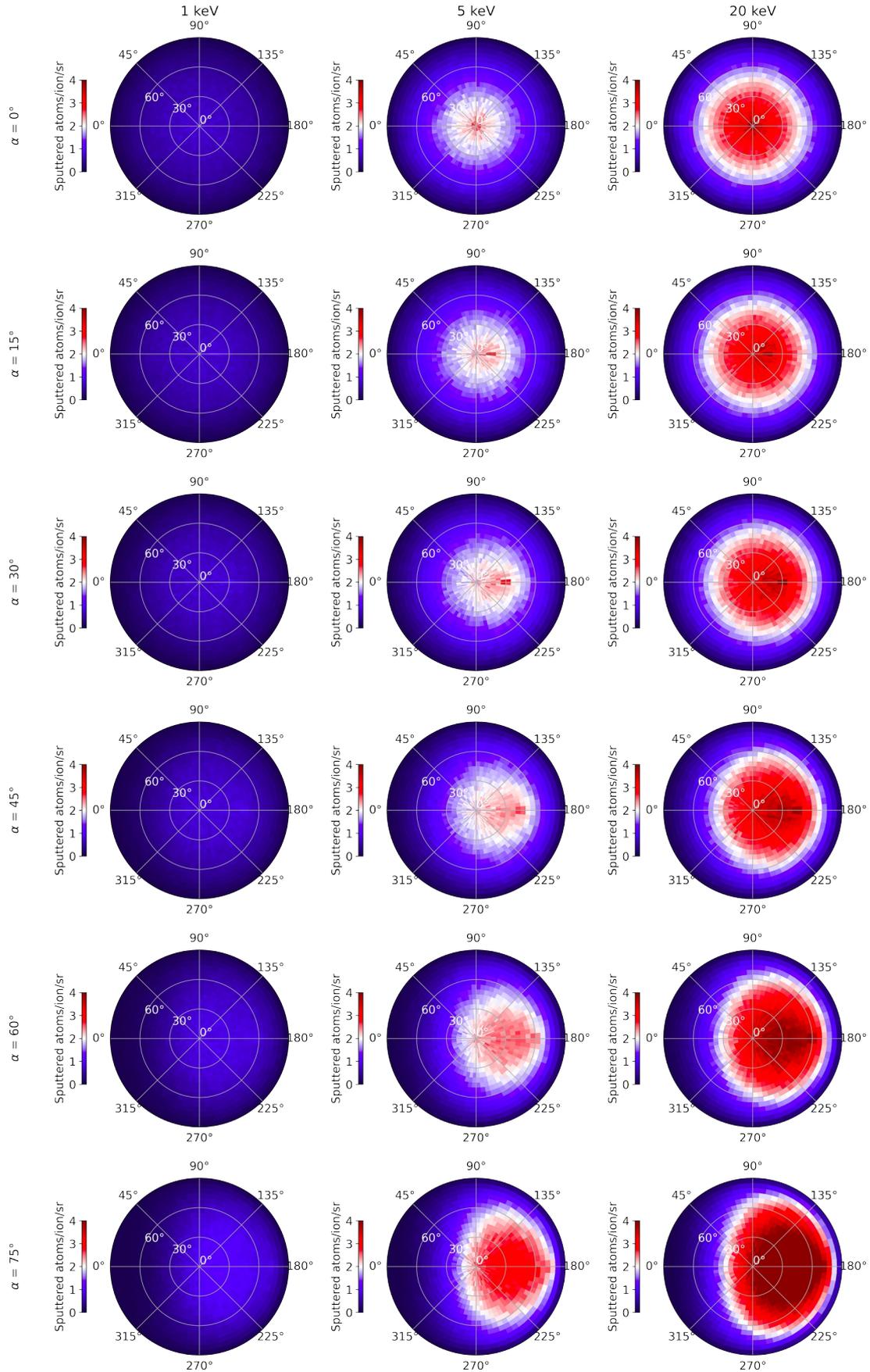

**Fig. 4**. Escaping sputtering yield angular distribution for a Kr⁺ ion beam on a Cu powder as a function of the beam energy and of the incidence angle. The ion beam is directed along $\varphi = 0°$ with an incident polar angle of $\alpha$.

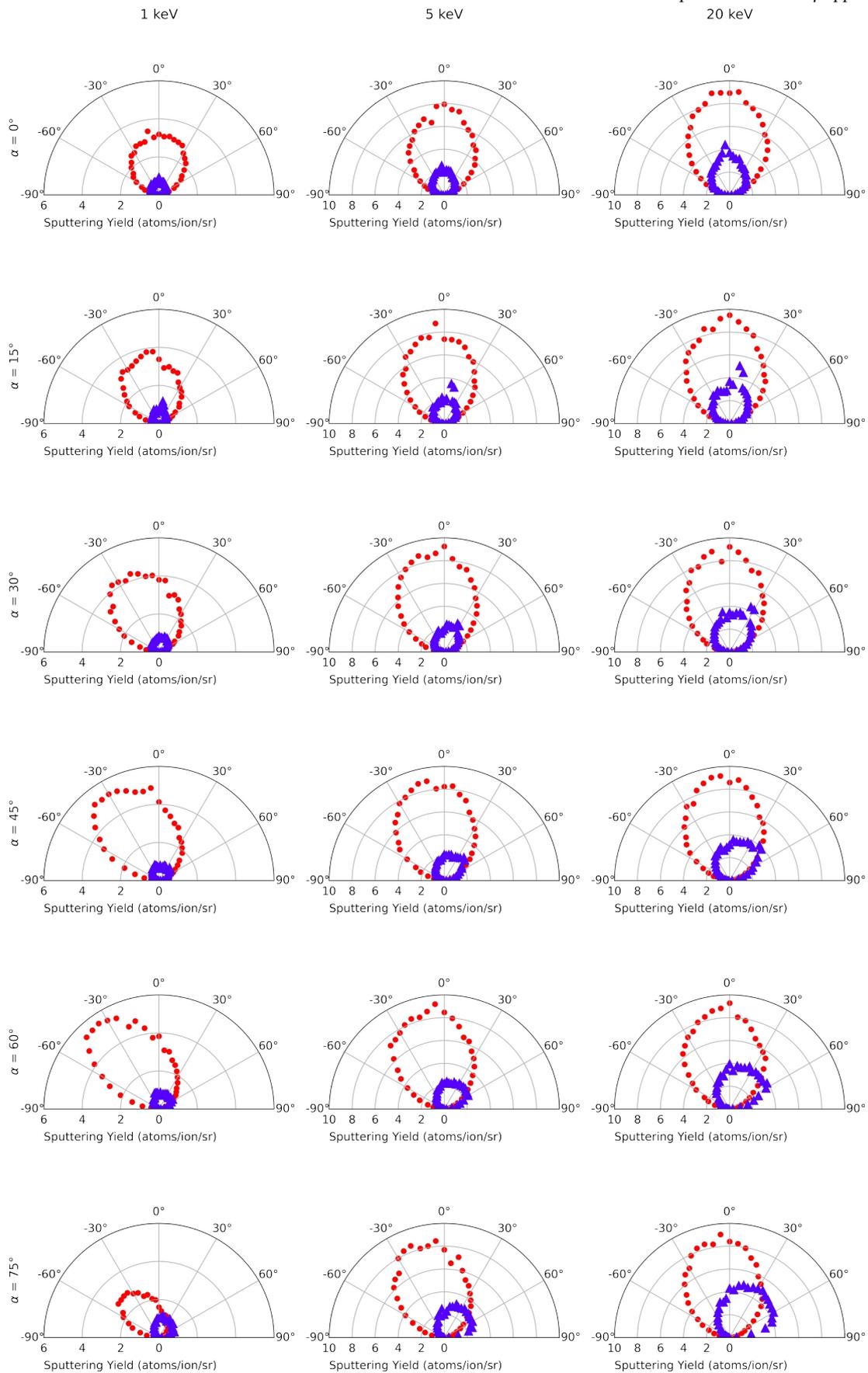

**Fig. 5.** Escaping sputtering yield in the incidence plan for the flat slab (red dots) and the powder (blue triangles) as a function of the polar angle for the different energies and incidence angles

considered here. The ion beam is incident from the positive polar angle quadrant. Note the change in sputtering yield scale between the left and middle columns.

| | | | 1 keV | | | | |
|---|---|---|---|---|---|---|---|
| α (°) | $Y_E$ (atom/ion) | $Y_R$ (atom/ion) | Escape (%) | $R_E$ | $Y_F$ (atom/ion) | $R_F$ | $Y_E/Y_F$ |
| 0  | 2.52 | 3.12 | 44.7 | 1.00 | 4.18 | 1.00 | 0.60 |
| 15 | 2.53 | 3.09 | 45.0 | 1.11 | 4.51 | 0.80 | 0.56 |
| 30 | 2.57 | 3.04 | 45.8 | 1.23 | 5.32 | 0.67 | 0.48 |
| 45 | 2.64 | 2.93 | 47.5 | 1.43 | 6.29 | 0.58 | 0.42 |
| 60 | 2.81 | 2.70 | 51.0 | 1.78 | 6.35 | 0.51 | 0.44 |
| 75 | 3.27 | 2.15 | 60.3 | 2.33 | 3.53 | 0.46 | 0.93 |

| | | | 5 keV | | | | |
|---|---|---|---|---|---|---|---|
| α (°) | $Y_E$ (atom/ion) | $Y_R$ (atom/ion) | Escape (%) | $R_E$ | $Y_F$ (atom/ion) | $R_F$ | $Y_E/Y_F$ |
| 0  | 6.46 | 7.52 | 46.2 | 1.01 | 8.32 | 1.00 | 0.78 |
| 15 | 6.51 | 7.49 | 46.5 | 1.16 | 9.02 | 0.90 | 0.72 |
| 30 | 6.61 | 7.42 | 47.1 | 1.35 | 11.1 | 0.85 | 0.60 |
| 45 | 6.80 | 7.13 | 48.8 | 1.62 | 14.6 | 0.81 | 0.47 |
| 60 | 7.27 | 6.55 | 52.6 | 2.02 | 17.9 | 0.77 | 0.41 |
| 75 | 8.5  | 5.21 | 62.0 | 2.60 | 15.5 | 0.70 | 0.55 |

| | | | 20 keV | | | | |
|---|---|---|---|---|---|---|---|
| α (°) | $Y_E$ (atom/ion) | $Y_R$ (atom/ion) | Escape (%) | $R_E$ | $Y_F$ (atom/ion) | $R_F$ | $Y_E/Y_F$ |
| 0  | 9.81 | 12.0 | 44.9 | 1.01 | 10.6 | 1.00 | 0.93 |
| 15 | 9.96 | 11.9 | 45.5 | 1.14 | 11.6 | 0.95 | 0.86 |
| 30 | 10.2 | 11.6 | 46.7 | 1.32 | 14.7 | 0.92 | 0.69 |
| 45 | 10.6 | 11.2 | 48.7 | 1.57 | 20.7 | 0.90 | 0.51 |
| 60 | 11.4 | 10.3 | 52.6 | 1.90 | 29.8 | 0.89 | 0.38 |
| 75 | 13.6 | 8.00 | 63.1 | 2.51 | 34.9 | 0.86 | 0.39 |

**Table 1.** Simulated escaping yield ($Y_E$), retained yield ($Y_R$), escape percentage, directionality ratio for the powder ($R_E$), sputtering yield from a flat sample ($Y_F$), directionality ratio for the flat slab ($R_F$), and the yield ratio $Y_E/Y_F$. The simulated results are shown for all three Kr$^+$ energies and six incidence angles α.

4. Discussion

4.1 Predominance of Backward Directed Ejecta

For the powder target, the backward directed fraction of the escaping yield increases with increasing $\alpha$. This is opposite to the behavior of the flat slab target where the forward directed fraction increases with increasing $\alpha$. The differences are quantified by $R_E$ and $R_F$. Both are 1 for $\alpha = 0°$, due to the cylindrical symmetry of the impact; but with increasing $\alpha$ at all energies considered here, $R_E$ increases whereas $R_F$ decreases.

This behavior of the powder can be attributed to the granular nature of the sample forming a porous dendritic or reticulated (i.e., interconnected) structure. The porosity of the powder causes a preferred direction for the escaping sputtered atoms (see Fig. 6).  Ions can penetrate through the interconnected voids of the powder and cause sputtering from underlying grains. Atoms sputtered from these underlying grains are most likely to escape from the powder if they are ejected roughly in the direction of the incident ions; the direction along which they are least likely to be shadowed by surrounding grains. The porosity effect is discussed in more detail in Section 4.2. In addition, for an ion beam irradiating the powder at a normal incidence angle, the ions are evenly distributed across the exposed surfaces of the impacted grains, resulting in a uniform angular distribution of the yield. However, as $\alpha$ increases, the ion-irradiated surface area of each grain becomes increasingly oriented in the backward direction, leading to shadowing of ejecta emitted in the forward direction.

Similar behavior has been seen with ENA backscattering from the lunar regolith surface. The Chandrayaan-1 Energetic Neutrals Analyzer (CENA; Kazama et al. 2007) observed the ENA reflected by the lunar regolith as a function of the solar zenith angle (SZA). The SZA is the angle between the Sun and the bulk normal to the lunar surface. Hence, the SZA of the impacting solar wind protons is equivalent to $\alpha$. ENA were observed to be predominantly backscattered toward the Sun for SZA $\geq 0°$ (Schaufelberger et al. 2011). This behavior is also observed in numerical simulations involving ENA backscattering from the lunar regolith surface (Szabo et al. 2023, Verkercke et al. 2023). These studies concluded that the ENA backscattering was caused by the granular nature for the regolith.

We are unaware of any previously published theoretical or experimental ejecta distribution results for powders that we could compare to.  However, enhanced backward directed ejecta is something that has been seen in some published works presenting combined theoretical and experimental sputtering results for a porous solid (Stadlmayr et al. 2020) and for a rough surface (Cupak et al. 2021).  Other studies, though, did not find any enhancement in the backward directed ejecta for rough surfaces of thin films (Biber et al. 2022) or pressed pellets (Biber et al. 2022, Brötzner et al. 2024, 2025)

Stadlmayr et al. (2020) and Cupak et al. (2021) both reported incidence plane sputtering yields for 2 keV $Ar^+$ impacting W targets with an incident angle of 60° relative to the bulk surface normal. Stadlmayr et al. (2020) studied flat and fuzzy W surfaces. Using a scanning electron microscope (SEM), they estimated that the W fuzz had a porosity of ~ 50% and a string or tendril thickness of ~ 40 nm. They did not report the root-mean-square (RMS) surface roughness or the mean value of the surface inclination angle, $\delta_m$. They found that the ejecta was forward directed for the flat surface, as expected at low impact energies where primary knock-on sputtering is important. For the W fuzz, they found that the escaping sputtered atoms were preferentially backward directed.

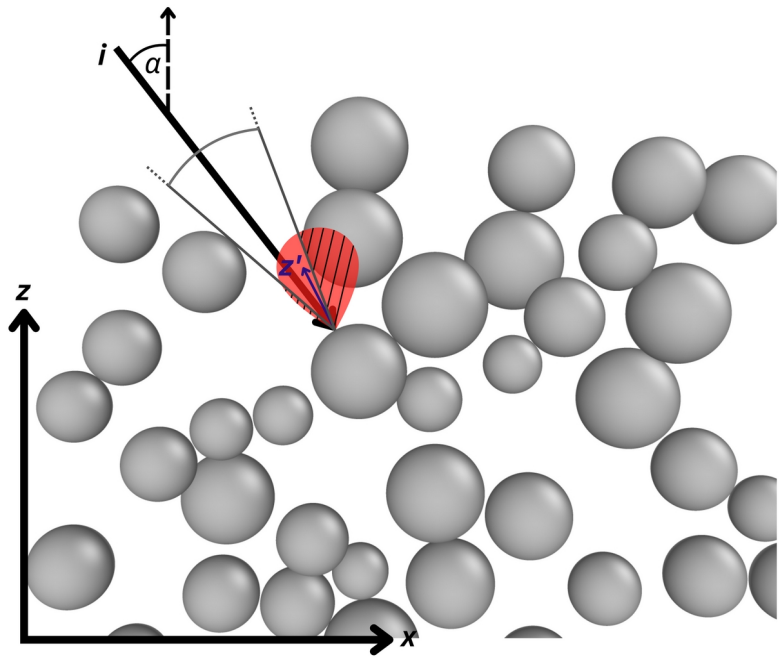

**Fig. 6**. A 50 µm thick slice of the powder in the incidence plane showing a single impact of an ion following the vector *i*, with an incidence angle $\alpha$. The ejecta distribution is represented by the red lobe around the local normal to the impact point *z'*. The dashed part of the ejecta distribution is shadowed while the undashed part is escaping. The escaping fraction of the ejecta distribution is contained in the grey angle, pointing toward the ion source.

Cupak et al. (2021) studied rough W surfaces, characterized by the RMS surface roughness and $\delta_m$. They were able to study sputtering as (RMS, $\delta_m$) evolved from ~ (1.7 nm, 7.6°) to ~ (110 nm, 15°) and lastly to ~ (1.5 µm, 36.5°). For the first two pairs, the ejecta was forward directed, indicating that primary knock-on sputtering was important. For the last pair, the ejecta showed a small enhancement in the backward direction, but nowhere nearly as extreme as found in our results or those of Stadlmayr et al. (2020).

Those findings, though, are to be contrasted with subsequent theoretical and experimental work on rough surfaces of thin films and pressed pellets, which have shown that the ejecta is primarily forward directed. Biber et al. (2022) studied 4 keV He$^+$ and 2 keV Ar$^+$ impacting at polar angles of 45° and 60° onto thin films generated by pulsed laser deposition from the enstatite and wollastonite ($\delta_m \approx$ 10° and 13°, respectively) and also onto pressed powder pellets formed from the same minerals ($\delta_m \approx$ 36° and 17°, respectively). Brötzner et al. (2024, 2025) studied 1 keV/amu H$^+$ and He$^+$ impacting a thin film of laser ablated lunar soil and a pressed pellet of the same lunar soil ($\delta_m \approx$ 6° and 28°, respectively). The RMS in these works was << 1 µm for the thin films and ~ 1 µm for the pressed pellets.

Combining our powder results with the work of Stadlmayr et al. (2020) and Cupak et al. (2021) suggests that porosity is the dominant contributor to the predominance of the backward directed ejecta for a loose powder. The porosity was ~ 50% for our work and estimated at ~ 50% for that of Stadlmayr et al., and qualitatively similar results were found despite the over-three-orders-of-magnitude difference in "granular" sizes: 50 – 90 µm diameter grains in our work versus the 40 nm string thickness in Stadlmayr et al. Earlier, we have defined this as granular roughness. But we could also consider these dimensions to be a "surface roughness" and compare to the simulations of Cupak et al. (2021), who found that the ejecta was forward directed for RMS roughnesses $\lesssim$ 110 nm. Yet, Stadlmayr et al. found the ejecta was predominantly backward directed for an RMS ~ 40 nm. We find similar results for an RMS of ~ 70 µm. Taken together, these imply that surface RMS cannot explain the predominance of the backward directed ejecta seen in these two simulations of porous materials.

The mean surface inclination angle is also unlikely to be able to explain this behavior. Biber et al. (2022) and Brötzner et al. (2024, 2025) find primarily forward directed ejecta for values of as large as $\delta_m \approx$ 36° and 28°, respectively. Cupak et al. (2021) found only a slight enhancement in the backward direction for $\delta_m \approx$ 36.5°. Stadlmayr et al. did not report a value for $\delta_m$. Our value of $\delta_m \approx$ 45° (see Section 4.5) is only slightly larger than those of Cupak et al. and Biber et al.; yet, we find a preponderance of ejecta in the backward direction. We attribute this finding to the interconnected porosity of our target leading to a preferential direction for the ejecta to escape from the powder.

To further investigate the effect of porosity, we ran our model for a closely-packed powder of $p =$ 0.32. This closely-packed powder is an extreme scenario of low porosity while still remaining a granular sample. We irradiated this sample with a 20 keV Kr$^+$ beam at $\alpha =$ 45° and 75°. Going from $p =$ 0.32 to 0.49, $R_E$ increased from 1.34 to 1.57 for $\alpha =$ 45°, and from 1.86 to 2.51 for $\alpha =$ 75°. Further increasing the porosity to $p =$ 0.62 and 0.87, our model predicts, respectively, $R_E =$ 1.62 and 1.69 for $\alpha =$ 45°, and $R_E =$ 2.52 and 3.01 for $\alpha =$ 75°. This shows that the escaping sputtered atoms from a low porosity powder exhibit a less pronounced backward-directed angular distribution compared to powders with higher porosities, and agrees with our hypothesis that the porosity is responsible for the predominance of backward directed angular distribution of the sputtered atoms. However, this dependency is not linear as shown by the large increase in $R_E$ between $p =$ 0.32 and 0.49 and the smaller increase going from 0.49 to 0.87.

4.2 Opposition Effect

The opposition effect is a well-known property in optical observations of airless planetary bodies such as asteroids (Gehrels 1956, 1957), the Moon (Gehrels 1964), and moons of other planets (Morrison et al. 1974). It occurs when a porous dendritic or reticulated surface structure is illuminated by light with wavelengths much smaller than the particles making up the surface. The effect is a direct result of the direction of the incident light being a preferred direction for the reflected light and occurs for phase angles between the incident and reflected light of $\lesssim 5°$ and results in a "surge" in reflected light for these angles (Hapke 1963). We find that a similar process occurs for ion sputtering from a powder. Ions can travel through the voids of the powder and impact underlying grains. Those atoms that are sputtered directly back toward the ion beam source can escape without being shadowed by overlying grains. The same will be true for atoms sputtered into a small cone around the direction of the incident ions; but atoms sputtered outside of this cone will be increasingly shadowed by the overlying grains, attenuating the escaping sputtering yield.

In our powder simulations, the opposition effect can be clearly seen in Figs. 4 and 5 for incident angles of $\alpha < 60°$ and phases angle of $|\theta - \alpha| \lesssim 5°$. Though the results are binned in 5° increments in $\theta$ and $\varphi$, this is still sufficient for being able to identify the effect. The effect fades for $\alpha > 60°$ because the surface becomes artificially less porous from the perspective of the ions, as the grains increasingly shadow each other, masking the voids in the powder and making it appear to be non-porous. It is worth noting at this point that it is unlikely that the ion sputtering field will be able to experimentally observe an opposition effect in the near future due to the laboratory challenges of mounting a detector within a 5° cone centered around the incident ion beam.

Interestingly, no opposition effect was noted in the theoretical calculations for porous solid samples (Stadlmayr et al. 2020), for non-porous rough surfaces (Cupak et al. 2021, Biber et al. 2022, Brötzner et al. 2024, 2025) or for pressed pellets (Biber et al. 2022, Brötzner et al. 2024, 2025). The difference compared to Stadlmayr et al. likely arises from the foam-like structure of the W fuzz. Hapke (1963) showed that the opposition effect can only be observed from a porous surface where the pores are interconnected (i.e., dendritic or reticulated), and not from a foam-like surface with disconnected pores. The difference compared to Cupak et al. (2021), Biber et al. (2022), and Brötzner et al. (2024, 2025) likely arises from the lack of porosity in their samples. Similarly, for our closely-packed powder sputtering simulations, the powder appears as a rough surface when $\alpha > 0°$ and we do not see an opposition effect.

The peak of the angular distribution pointing in the direction of the ion-beam origin was also observed in ENA backscattering studies. The Chandrayaan-1 CENA instrument observed ENA from a range of viewing angles during its operation around the Moon. CENA detected an enhancement of the number of ENA reflected in the direction of the Sun (Schaufelberger et al. 2011). Unfortunately, the observations were averaged using 15° SZA bins, which does not enable one to say definitely if there was an opposition effect. However, it does support that hypothesis. This backscattering effect was attributed to the porosity of the lunar regolith and was explained by numerical simulation that considered the reflection ENA reflection of regolith-like structures (Verkercke et al. 2023).

4.3 Total Escaping Sputtering Percentage and Yield

Over the energy range considered, the escape percentage increases from ~ 45% for $\alpha = 0°$ to ~ 62% for $\alpha = 75°$. The percentage retained correspondingly decreases from ~ 55% to ~ 38%. We find that these percentages are nearly independent of ion energy, indicating that they are primarily influenced by the granular structure of the powder. For $\alpha = 0°$, ions that impact the top of a grain do so at a normal local incidence, producing the lowest yield but highest escape probability. As the impact point moves toward the edge of a grain, the local incidence angle increases, increasing the yield while decreasing the escape probability, due to the ejecta lobe for these local incidence angles being pitched in the forward direction which is further into the powder. As $\alpha$ increases, ions that impact the top of a grain hit with a local incidence angle equal to $\alpha$ (Szabo et al. 2022a, 2023, Verkercke et al. 2023). The increasing $\alpha$ leads to an increase in the yield and thus an increased total escape percentage. Additionally, for $\alpha = 0°$, ions penetrate deeper into the powder before impacting a grain. A deeper impact point results in atom ejection occurring from deeper within the powder (Verkercke et al. 2023), thus increasing the likelihood of shadowing. However, with increasing $\alpha$, the maximum penetration depth of an ion into the powder decreases, leading to a decrease in the shadowing probability. The combination of the impact angle and correspondingly reduced shadowing produce the observed slow increase in the escape percentage. Similar behavior is also observed in the reflection rate of energetic neutral atoms (ENA) from lunar regolith, showing the wide range of applicability of such simulations (Szabo et al. 2022a, Verkercke et al. 2023).

Our results indicate that there is a higher escape percentage from a powder than that predicted by Cassidy and Johnson (2005) for a planetary regolith. For $\alpha = 0°$, our escape percentage of ~ 45% is 5.5 times larger than the 8% of Cassidy and Johnson for their forward-directed sputtering yield distribution (i.e., primary knock-on sputtering) and almost 2 times larger than their 27% for their cosine-law angular yield distributions (i.e., secondary knock-on sputtering). We attribute their under-prediction of the escape percentage to their use of the approximate yield formula for the total sputtering yield from a flat surface of

$$Y_F(\alpha)/Y_F(0) \approx \cos^{-n}(\alpha), \qquad (2)$$

where $n \geq 1$. For local incidence angles $\alpha > 50°$, this led to yields far greater than those that we obtain using SDTrimSP. Most of this overestimated sputtering appears to be retained by the regolith in their model, thereby leading to an underestimate in the escape percentage. Cassidy and Johnson also found that their results were largely independent of the porosity and grain size of the regolith. In Section 4.5, we show that the results do indeed depend on porosity. In a future work, we will explore the role of grain size, though similar work on ENAs finds that the grain size distribution has little influence on the ENA reflection fraction (Verkercke et al. 2023).

We can also compare to the predicted $Y_E/Y_F$ ratio of Cassidy and Johnson (2005) for $\alpha = 0°$, which is given in their work as $Y_R/Y_F$. They present results for $Y_E/Y_F$ for forward-directed ejecta and for a cosine emission distribution law, the former of which is specular and depends on the incident angle $\alpha$ of the incoming ion and the latter of which depends on the emission angle $\theta$ of the ejected atom as $f(\cos(\theta)) \approx 2\cos(\theta)$. These two extreme cases indirectly represent the energy of the incident ion with the forward-directed being appropriate for a low energy (where primary knock-on collisions dominate) and the cosine emission law for a high energy (where secondary knock-on collisions dominate). Cassidy and Johnson predict for forward-directed ejecta a ratio of $Y_E/Y_F = 0.19$ for their assumed $n = 1.6$. For our 1 keV results, which are closest to this case, fitting $Y_F(\alpha)$ for $\alpha \lesssim 50°$, we find $n = 1$ and $Y_E/Y_F = 0.60$. The results of Cassidy and Johnson are significant smaller than ours. For our 20 keV results, which are the closest to the cosine emission distribution law, fitting $Y_F(\alpha)$ for $\alpha \lesssim 50°$, we find $n = 1.4$ and $Y_E/Y_F = 0.93$. Again, the results of Cassidy and Johnson are significantly smaller with values of 0.45 and 0.64 for $n = 1.0$ and 1.6, respectively. We attribute the systematic underestimates of Cassidy and Johnson to their overestimate of the sputtering yield for large values of $\alpha$, as discussed earlier. Interestingly, we both find a similar energy dependence for the ratio increasing as the sputtering evolves from low to high energies.

Our modeled powder results for the yield escaping the sample can also be compared to published theoretical and experimental sputtering results for porous solids (Stadlmayr et al. 2020) as well as for non-porous rough surfaces of both thin films and pressed powder pellets (Cupak et al. 2021, Biber et al. 2022, Brötzner et al. 2024, 2025). These are not exact analogs for a loose powder; so, much of the comparison is qualitative in nature, in part to point out the complexity of ion sputtering and the importance of properly accounting for the impactor and target properties.

Stadlmayr et al. (2020) studied sputtering by 2 keV $Ar^+$ impacting flat and fuzzy W surfaces with an incidence angle of 60° relative to the bulk surface normal. They used a quartz crystal microbalance (QCM) as a catcher and measured the mass gain from the ejecta as they moved the collector along the incidence plane. For a QCM position approximately normal to the target ($\theta \approx 0°$; their $\Delta x = 6$ mm), they found that their "catcher yield" for the W fuzz was ~ 17 times lower than that for a flat surface. A similar reduction was also observed by Nishijiama et al. (2011) for their primarily $Ar^+$ plasma sputtering a W fuzz and flat surface at energies of $E_i = 50 - 110$ eV. They reported a reduction in the total sputtering yield by a factor 10. Comparing first to the catcher yield results of Stadlmayr et al., for our $\alpha = 60°$ and $E_i = 1$ and 5 keV results we find a maximum reduction of only a factor of ~ 4 for the sputtering yield in the incidence plane when summed over the same ejecta polar angles (Fig. 5). This factor decreases to ~ 3 for $E_i = 20$ keV. Comparing to the total sputtering yield results of Nishijiama et al., we find a maximum reduction in our total sputtering yield of ~ 2.6 (Table 1). In neither case do we see an order-of-magnitude decrease. We attribute these differences to two different effects: (1) the higher porosity of the W fuzz compared to our Cu powder, and (2) to the potential oxidation of the samples used in the experiments. Nishijiama et al. report a porosity of $p \sim 0.90 - 0.95$, which they calculated using SEM images to determine the volume of the fuzz combined with the mass change of the target before and after the fuzz was scraped off. Stadlmayr et al. also used SEM images to determine their porosity, reporting a value of $p \sim 0.5$. This is much smaller than the ~ 0.9 reported by others for W fuzz (Baldwin and Doerner 2010, Nishijiama et al. 2011) and is likely due to the inability of SEM images to detect any nanometer-scale porosity of the fuzz strings. In future modeling studies, we will explore the effects of these high foam-like porosities on the nanometer scale. Studies of such finely resolved structures are readily amenable to the approach that we have developed here but are extremely challenging for voxel methods such as SDTrimSP-3D and TRI3DYN, which would require a computationally prohibitive number of voxels to resolve the nanometer-scale structure (e.g., Stadlmayr et al. 2020). Regarding the oxidation of the sample, Kupart et al. (2010) showed that oxidized metals exhibit larger surface binding energies (SBEs) than in their pure form equivalent. Increasing the SBE has been shown theoretically and experimentally to reduce sputtering yields (Oechsner 1975). This implies that even a partial oxidation of the respective samples used by Nishijiama et al. and Stadlmayr et al. could have caused a reduction of their yield results.

Moving on to non-porous rough surfaces of thin films, Cupak et al. (2021) theoretically and experimentally studied sputtering by 2 keV $Ar^+$ ions impacting W-coated quartz crystals. The roughness of the physical-vapor-deposited W was controlled using quartz crystal substrates with different surface finishes: polished versus unpolished. They found for a rough surface, that for any incidence angle the total yield measured was either equal to or smaller by up to a factor of ~ 1.3 than the yield measured from a flat sample. The yield dependence versus the incidence angle was similar for both the flat and the rough samples, with an increase of the yield up to $\alpha_{peak} \approx 60° - 65°$, above which the yield decreased. At their energy used, primary knock-on collisions are important and so we compare their results to our 1 keV results. While we also find that the theoretical yield from a powder is lower than for a flat surface (cf., Fig. 7), we see a significant difference in the behavior of the yield versus the incidence angle, with the powder exhibiting a sputtering yield that continually increases with increasing incidence angle and for $\alpha \approx 85°$ still shows no sign of turning over.

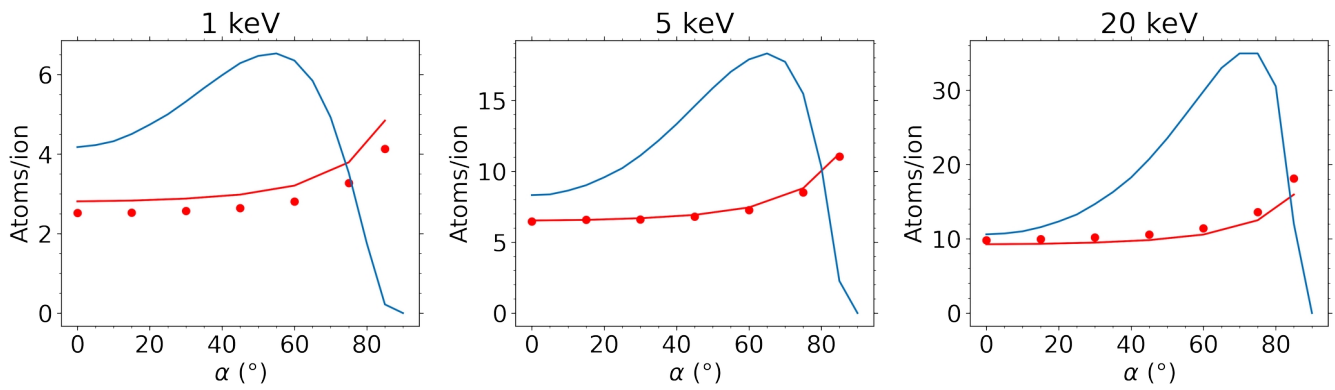

**Fig. 7.** Total sputtering yield as a function of the ion beam incidence angle for the flat surface (solid blue line) and for the powder (red dots) described in the text. Equation (6) is plotted as the red solid line for $\alpha'_M = 45°$.

Biber et al. (2022) theoretically and experimentally studied 4 keV He$^+$ and 2 keV Ar$^+$ impacting at polar angles of 45° and 60° onto thin films generated by pulsed laser deposition from the enstatite and wollastonite and also onto pressed powder pellets formed from the same minerals. The pressed pellets presented, respectively, $\delta_m$ of ≈ 36° and ≈ 17°. Comparing to their results for a flat surface, for the largest $\delta_m$ they observed an increase of the yield by the pellet for small incidence angles, and a reduction of the yield for larger incidence angles. This contrasts with the findings of Cupak et al. (2021) for rough surfaces and our results for powders, both studies of which find total yields are always smaller than that of a flat surface. However, Biber et al. do observe a change in the yield dependency between a rough and a flat surface that is similar to the one we observe between a powder and a flat surface.

Brötzner et al. (2024, 2025) experimentally and theoretically studied sputtering by 1 keV H$^+$ and 4 keV He$^+$ impacting a flat surface ($\delta_m$ ≈ 6°) and a pressed pellet of lunar regolith sample 68501 from Apollo 16 ($\delta_m$ ≈ 28°). Additionally, they numerically studied sputtering by the same ions impacting a simulated regolith powder of $p$ = 0.8. Similar to Biber et al. (2022), but unlike Cupak et al. (2021), they find that the sputtering yield from the rough surface compared to a flat surface is larger for small impact angles and smaller for large impact angles. They also showed that, for the same material composition, the escaping yield of a rough surface is always greater than the yield escaping from a loose powder. We attribute this to the lack of porosity of the rough surface compared to the powder. Additionally, the incident angular dependence of the yield that they report for the rough surface and regolith are similar to what we find for a powder,

Perhaps the one set of published results that are most appropriate to compare to our results are the regolith simulations of Brötzner et al. (2024, 2025). They report their mass yield in amu/ion and assume that the stoichiometry of the ejecta is identical to that of the target in order to convert to atoms/ion. They find a regolith yield for $\alpha$ below ~ 20° that is similar to that from a flat. Above ~ 20°, the yield from the flat and regolith both increase but the yield from the flat increases more rapidly than from the regolith, until at $\alpha_{peak}$ ≈ 80° the yield from the flat turns downward while that from the regolith continues to increase. The results of Brötzner et al. were for ion energies where primary knock-on sputtering is important, whereas our results range from 1 keV where primary knock-on sputtering is important to 20 keV where secondary knock-on sputtering is important. We find that $\alpha_{peak}$ increases with impact energy but never quite reaches the ≈ 80° of Brötzner et al. In addition, we find that the escaping yield increases with impact energy, but it is only at the highest energy investigated of 20 keV that the regolith yield is nearly equal to that of the flat for $\alpha \lesssim 20°$. It is difficult to make a more quantitative comparison than this because of the difference between the impactors, energies, target compositions, surface binding energy assumptions, and porosities.

Lastly, our simulations show that $Y_E$ increases with increasing energy and incident angle of the Kr$^+$ impactors. However, for any energy, $Y_E/Y_F$ first decreases with increasing incidence angle, and reaches a minimum around $\alpha$ ~ 45-60°, as shown in Table 1. Above this incidence angle, $Y_E/Y_F$ increases again as $Y_F$ starts decreasing but $Y_E$ keeps increasing with $\alpha$. Looking forward to the experimental work of Bu et al. (in preparation), we predict a ratio of $Y_E/Y_F$ = 0.51 for 20 keV Kr$^+$ impacting at $\alpha$ = 45° onto spherical Cu grains with a uniform grain size distribution between 50 and 90 µm.

## 4.4 Lack of Apparent Evolution from Primary to Secondary Knock-on Sputtering

Building on the discussion in Sections 4.1 and 4.2, it is easy to explain the lack of evolution of the sputtered atoms angular distribution when considering a powder (Fig. 4). The directionality of the sputtering yield from a flat surface is characterized by $R_F \leq 1$ for any incidence energy for $\alpha > 0°$, indicating a preferential forward-scattering of the ejected atoms. Moreover, $R_F$ moves closer to 1 with increasing impact energies as the sputtering process moves from being dominated by primary knock-on to being dominated by secondary knock-on collisions, the latter of which produces cylindrical symmetry in the ejecta. This cylindrical symmetry is not observed for the escaping yield from a powder, as shown by the Fig. 4. This is due to the interconnected porosity and geometry of the spherical grains. The former produces a preferred direction for the ejecta to escape the powder. The latter results in shadowing of forward directed ejecta. Together, these result in the ejecta distribution from a powder differing from that of a flat surface.

## 4.5 Potential Universal Scaling Laws

### 4.5.1 Escaping Yield of Powders

Guided by our result, we find that the escaping yield versus $\alpha$ for a powder can be well reproduce using

$$Y_E(\alpha,p) \approx E(\alpha,p) \cdot Y_F(\alpha'_M), \qquad (3)$$

where the escape fraction $E(\alpha,p)$ is a function of the incident angle $\alpha$ and porosity $p$; and $\alpha'_M$ is the mean local incidence angle on the individual grains, which is equivalent to the mean inclination angle of the sample $\delta_m$. This approximation is motivated, in part, by the suggestion that sputtering from a rough surface can be approximated by that from a flat surface for $\alpha = 45°$ (Küster et al. 1998, Wurz et al. 2007, 2022). As shown in Fig. 8, the most probable $\alpha'_M$ for an ion impacting a spherical grain is ~ 45°, making this interaction dominant for the total ejecta, independent of the global incidence angle.

We have found that the escape fraction depends primarily on the porosity of the powder $p$. This is similar to what has been found for the ENA backscattering efficiency, which also depends on the porosity of the regolith (Szabo et al. 2022, Verkercke et al. 2023). To verify our hypothesis, we have determined the escape fraction for the unshaken Cu powder, but where the top layer containing a fairy-castle structure was sliced off. We also determined the escape percentage of a powder with a close packing to test the limit of our prediction. The respective porosities of these two powders are 0.62 and 0.32.

We also selected two fairy-castle cases: the first is the unshaken and unsliced Cu powder and the second is a regolith-like powder similar to that presented in Verkercke et al. (2023). Because the incident ions are most likely to impact the fairy-castle structure before reaching the underlying bulk powder (Verkercke et al. 2023), we calculate an effective porosity $p_{eff}$ by averaging over the topmost few millimeters for each case. The resulting value of $p_{eff}$ are 0.78 and 0.87, respectively.

Using the four structures described above and including the shaken and sliced Cu powder of $p$ = 0.49 (see Fig. 9), we find the empirical relation:

$$E(0°,p) \approx -0.45\,p + 0.68 \qquad (4)$$

Equation (4) reproduces the model outputs to within 2% for all five cases considered. We can next determine the escape fraction versus $\alpha$ using the five cases and 19 angles considered, as shown in Fig. 9. We find that $E(\alpha,p)$ increases as $\alpha$ increases and can be approximated as:

$$E(\alpha,p) \approx E(0°,p) \cdot [\cos(\alpha)]^{-0.103/p} \qquad (5)$$

Equation (5) matches all the model outputs to within 7%. This formula works well for high porosities, but it fails for combinations of low $p$ and large $\alpha$, i.e., small $\cos(\alpha)$, where it can predict an unphysical escape fraction of greater than 1. For these combinations, the surface starts to appear more as a rough surface than a porous sample, which could explain why the relationship diverges from the numerical simulations.

Using Equations (4) and (5), we can re-write (3) as:

$$Y_E(\alpha,p) \approx (-0.45\,p + 0.68) \cdot [\cos(\alpha)]^{-0.103/p} \cdot Y_F(\alpha'_M) \qquad (6)$$

This last equation gives a relation between the total yield escaping from a powder of porosity $p$, for an ion beam impacting the sample with a bulk incidence angle $\alpha$. It further depends only on the yield from a flat surface, with a composition similar to that of the powder, at the mean local incidence angle $\alpha'_M$ (with $\alpha'_M \approx 45°$ for spherical grain). This equation is plotted in Fig. 7 as the solid red line, with $\alpha'_M$ taken as equal to 45°. Equation (5) reproduces the model outputs within ~ 15% for 1 keV $Kr^+$ and within ~ 5% for $E_i$ = 5 and 20 keV. Our work here suggests a universal scaling law that relates the total escaping sputtering yield from a powder to that of a slab. Clearly, though, further theoretical and experimental studies are needed to explore the effect of the ion type, ion energy, target composition, grain size distribution, grain shapes, porosity, etc. in order to verify the validity of this potential scaling law.

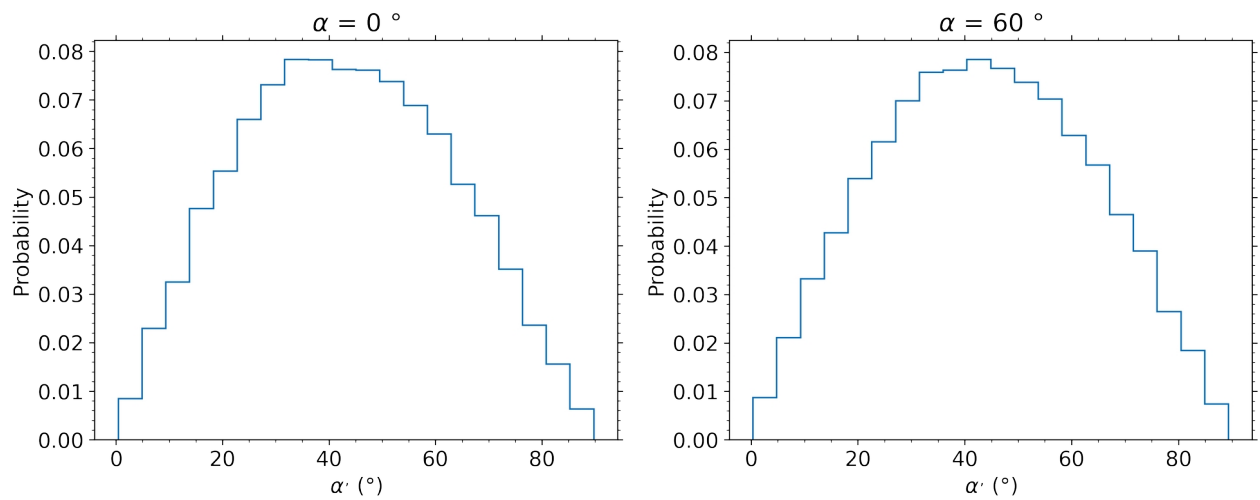

**Fig. 8.** Probability distributions of the local incidence angles experienced by ions launched at a Cu powder at $\alpha = 0°$ and $60°$. The corresponding mean incidence angles are $\alpha'_M = 43.3°$ and $43.9°$, respectively.

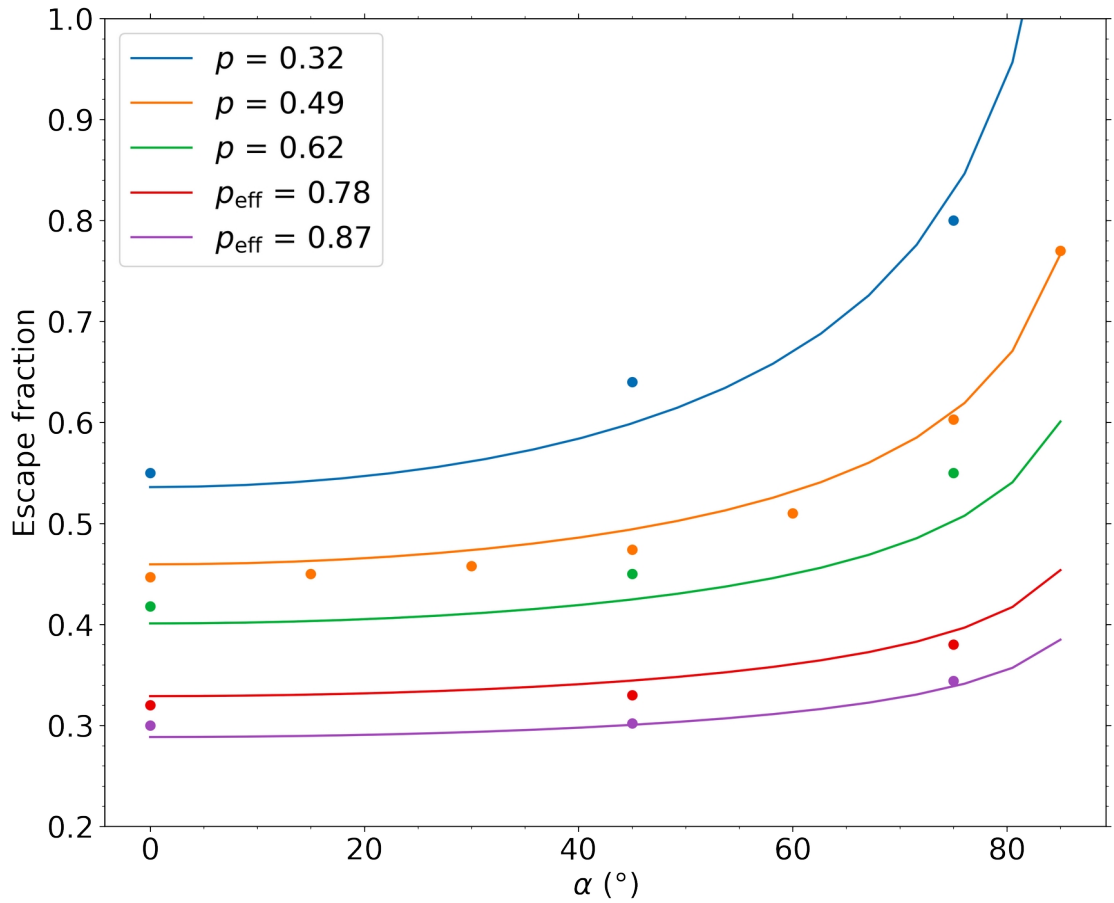

**Fig. 9.** Escape fraction of the yield out of a powder as a function of the incidence angle of the ions for the five structures studied. The dots are the values simulated by our model and the solid lines are given by Equation (5).

4.5.2 Reflection of ENAs

Our derivation of Equation (6) suggests that a similar relationship should exist for ions reflected as ENAs, namely

$$R(\alpha,p) \approx (-0.45\, p + 0.68) \cdot [\cos(\alpha)]^{-0.103/p} \cdot R_F(\alpha'_M). \quad (7)$$

Here, $R(\alpha,p)$ is the reflection fraction from a powder and $R_F$ the reflection fraction from a flat surface. Geometrical effects are accounted for by $p$ and $\alpha$, while the ion type and energy and the sample composition are accounted for by $R_F$.

As a test of this hypothesis, we here assume solar wind protons impacting the same two regolith structures as in Verkercke et al. (2023), each with a grain size of 50 µm. In that work, they showed that the grain size distribution has little influence on the ENA reflection fraction. One of the regolith samples considered has a porosity $p = 0.55$ and no fairy-castle structure. The second regolith sample considered has a fairy-castle structure with $p_{eff} = 0.87$, similarly to the case shown for the sputtering yield in Fig. 9. In each case $\alpha'_M \approx 45°$. The grains are considered to be pure silica. $R_F(\alpha'_M)$ is then the reflection coefficient of H on a flat $SiO_2$ sample, taken from the MD results of Verkercke et al. (2023).

Using Equation (7), the ENA reflection fraction as a function of SZA, which is equivalent to $\alpha$, is shown in Fig. 10. The dots show the output from the simulations of Verkercke et al. (2023). The solid lines show the results of Equation (7) for the two porosities considered and the MD calculated $R_F(45°) = 0.43$. Equation (7) reproduces the model outputs to within ~ 10%. The Chandrayaan-1 measurements of lunar ENA are also plotted with their uncertainty range (Vorburger et al. 2013). The empirical relation reproduces well the outputs of the simulation and is within in the range of the ENA observations.

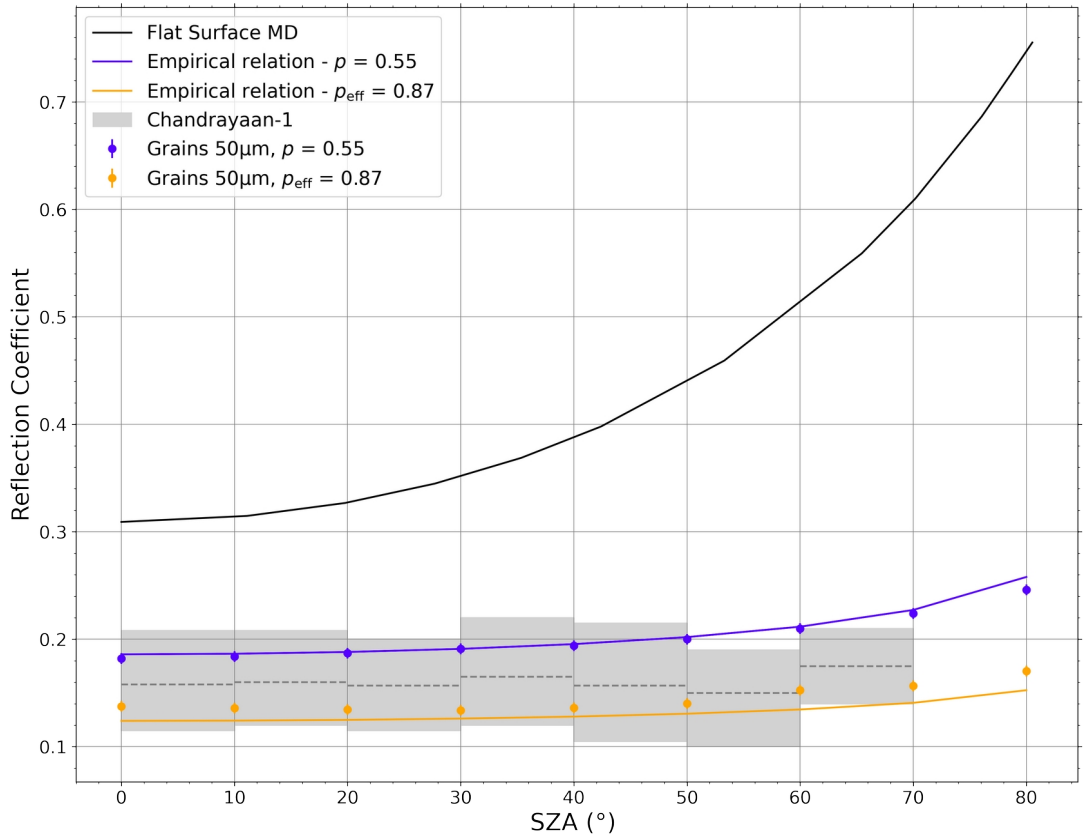

**Fig. 10.** Reflection fraction of H ENA as a function of the SZA. The black solid line shows the reflection of H on a SiO$_2$ film predicted by MD. The orange and blue dots present the simulations from Verkercke et al. (2023). The solid lines use Equation (7) for the corresponding porosities and value of $R_F(45°)$. The ENA coefficient measured by Chandrayaan-1 are shown by the dashed grey lines and their associated uncertainties by the grey error bands.

4.5.3. Escaping Sputtering Yield Angular Distribution of Powders

Based on our model, we propose a fitting function that can describe the angular distribution of the escaping sputtered atoms as a function of $\alpha$. Specifically, we have fitted the angular distributions of the escaping yields for the powders presented in Fig. 4, normalized so that the integral over all solid angle was equal to 1. These normalized distributions are shown in Fig. 11. We use a sum of real spherical harmonics (Atkinson & Han 2012) for the fit giving:

$$f(\varphi,\theta)=\sum_{l=0}^{\infty}\sum_{m=-l}^{l} f_{lm} Y_{lm}(\varphi,\theta) \qquad (8)$$

where $f_{lm}$ are fitting coefficients and $Y_{lm}$ are the spherical harmonic functions.

We first performed a fourth-order fit, up to $l = 4$, for each incidence angle and energy. This resulted in 25 different $f_{lm}$ coefficients for each angle and energy pair. We used the coefficient of determination $r^2$ to quantify the goodness of the fit. This value indicates the fraction of the variance that is explained by the fit, and varies between 0 and 1, with 1 being the perfect fit. The fourth-order fits all have correlation coefficients $r^2 \geq 0.93$, with mean absolute errors (MAE) < 8%. The fit does not reproduce the opposition effect, but the low MAE demonstrates that this introduces a small overall error. For modeling ease, we also explored using a smaller number of fitting parameters. For each angle and energy pair considered, while all 25 $f_{lm}$ coefficients are non-zero, the 9 coefficients with $m = 0$ or $m = 1$ are greater than the others by one or several orders of magnitude. Setting the $m < 0$ and $m > 1$ coefficients to zero, the resulting fits give $r^2 \geq 0.93$. Putting any of the dominant coefficients to zero results in a majority of the fits have an $r^2 \leq 0.85$ (MAE $\geq$ 15%), with some fits as low as $r^2 = 0$ (MAE > 100 %).

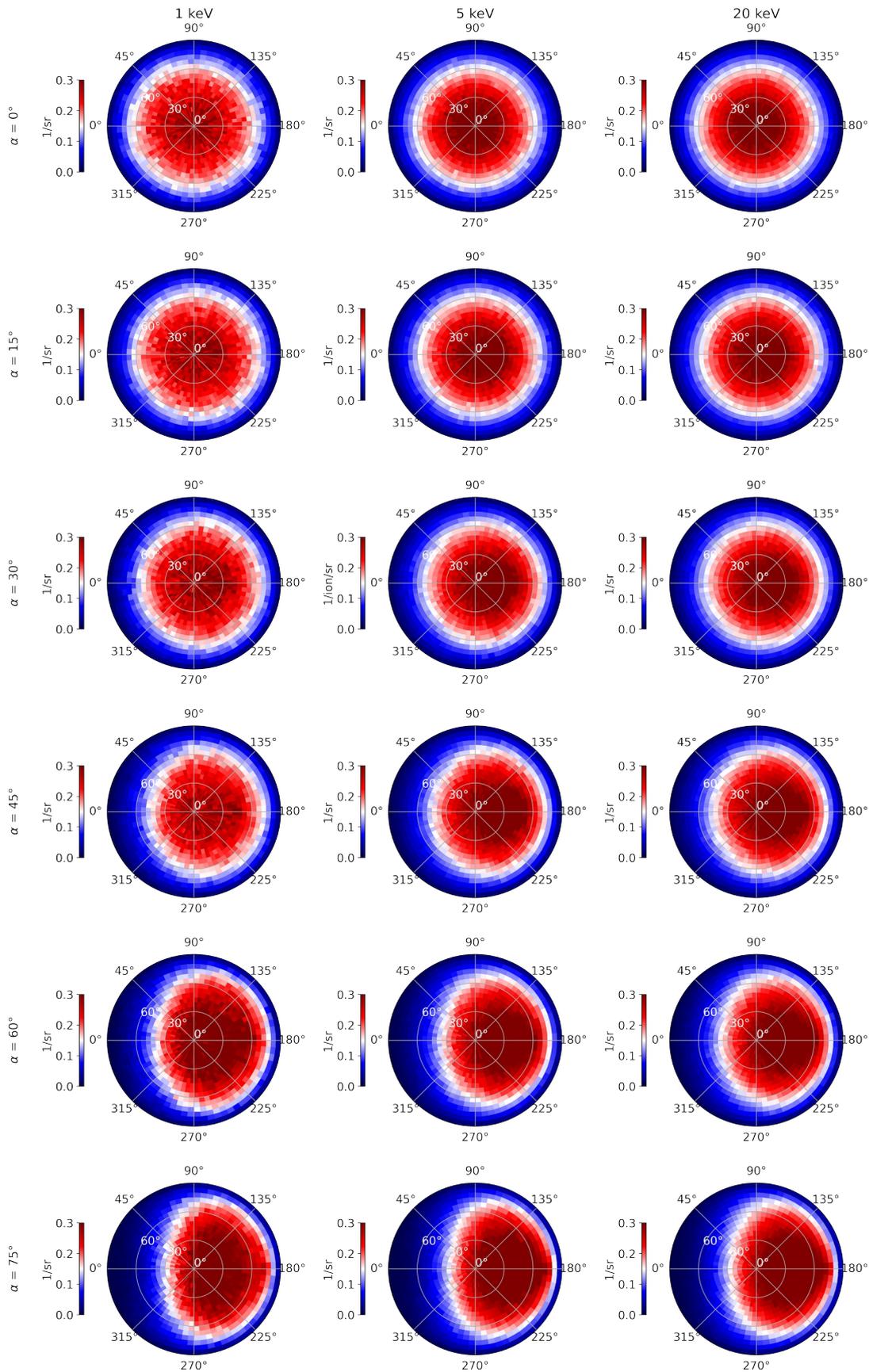

**Fig. 11**. Same as Fig. 4 but normalized by the total sputtering yield per ion ($Y_E$).

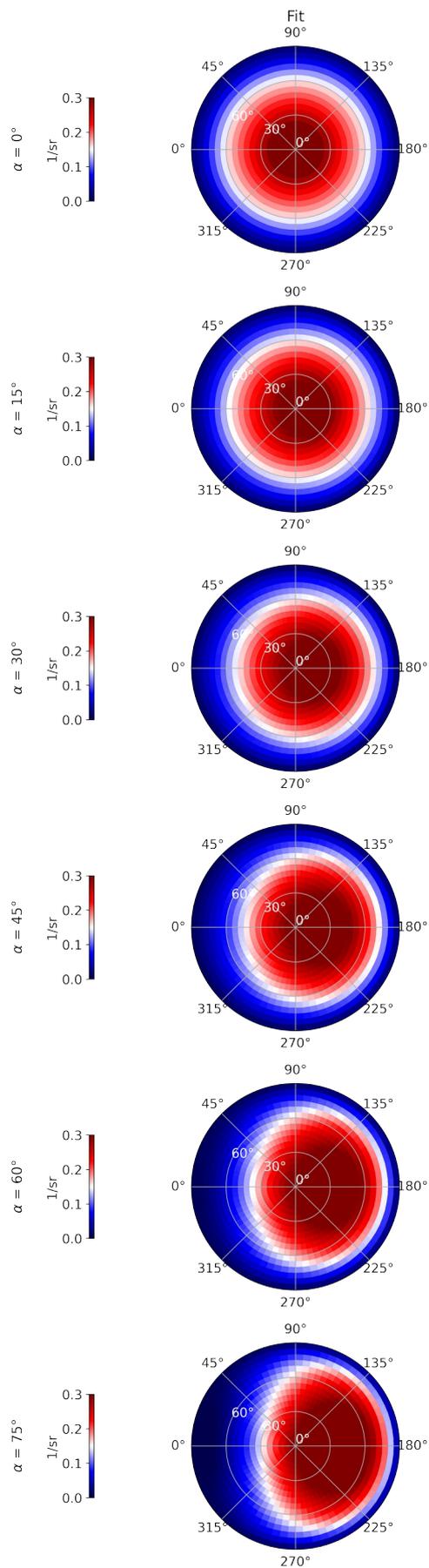

**Fig. 12**. Fit to the escaping sputtering yield angular distribution normalized by the total sputtering yield per ion for a $Kr^+$ ion beam on a Cu powder as a function of the incidence angle. The ion beam is directed along $\varphi = 0°$ with an incident polar angle of $\alpha$.

We can further simplify the fits to the angular distribution taking into account the approximately similar behavior for the dominant coefficients with impact energy. Fitting simultaneously the angular distribution of the three different impact energies studied for each $\alpha$ gives an $r^2 \geq 0.9$ and an MAE $\leq 10\%$. Fig. 12 shows this combined fit using only the dominant coefficients for the different incidence angles. The corresponding fit parameters are given in Table 2.

We have also explored the applicability of the derived fit for porosities different from $p = 0.49$. To study this, we use the fitting functions presented in Fig. 12 and Table 2 and analyze the variation of the $r^2$ as a function of porosity. We consider $p = 0.32$, 0.49, and 0.87. Fig. 13 shows the normalized angular distributions for these three cases at three different $\alpha$. Comparing the fit and the angular distribution of the yield for $p = 0.32$ gives $r^2$ values < 0.9 with an MAE ~ 15%, for $p = 0.49$ we find $r^2$ values $\geq 0.9$ with an MAE ~ 10%, and for $p = 0.87$ we have $R^2$ values $\geq 0.9$ and an MAE ~ 10%. This implies that the variation of the angular distribution is only marginal above a certain porosity.

We thus conclude that the proposed fit function with its set of dominant parameters can reproduce the angular distribution of the sputtered atoms at any energy between 1 keV and 20 keV, and for any porosity $p \geq 0.49$ with a determination coefficient $r^2 \geq 0.9$ and an MAE $\leq 10\%$ for any incidence angle between 0° and 75°. For $\alpha > 75°$, the fit has not been computed, but such large incidence angles are rarely observed naturally or laboratory measurements.

To conclude, we propose that the absolute doubly differential escaping sputtering yield can be given by the fit of Equation (8) scaled by the total yield per ion given by Equation (6). These functions could also be used in the study of airless bodies and their exosphere, as the angular distribution of the sputtered atoms play an important role in the global migration of atoms around these bodies (Teolis et al. 2023).

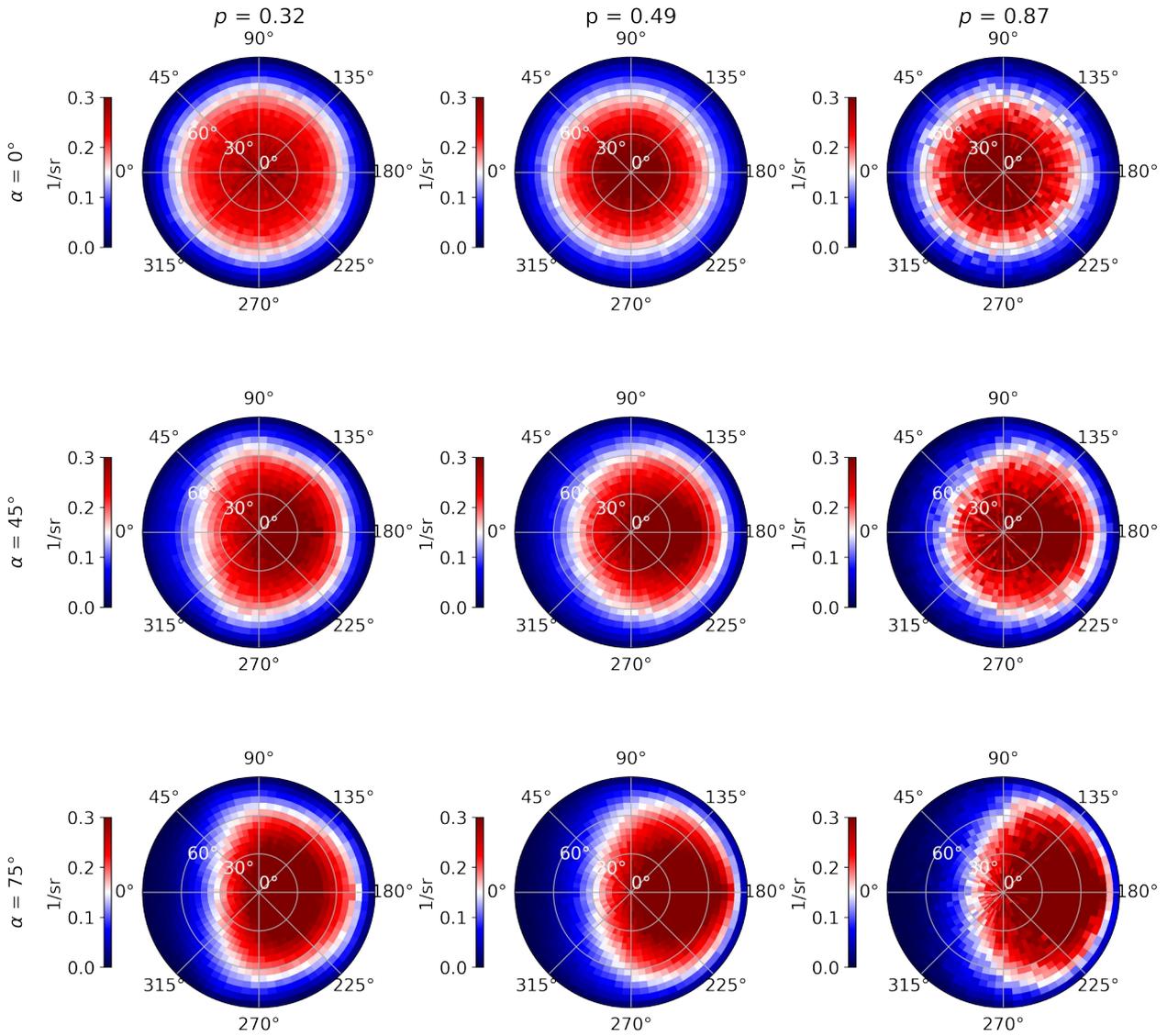

**Fig. 13**. Escaping sputtering yield angular distribution normalized by the total sputtering yield per ion ($Y_E$) for a Kr$^+$ ion beam on a Cu powder as a function of the powder porosity and of the incidence angle. The ion beam is directed along $\varphi = 0°$ with an incident polar angle of $\alpha$.

|  | $l = 0$ | $l = 1$ | | $l = 2$ | | $l = 3$ | | $l = 4$ | |
|---|---|---|---|---|---|---|---|---|---|
|  | $m = 0$ | $m = 0$ | $m = 1$ | $m = 0$ | $m = 1$ | $m = 0$ | $m = 1$ | $m = 0$ | $m = 1$ |
| **0°** | 3.259 | -3.756 | 0.078 | 3.272 | -0.106 | -1.537 | 0.072 | 0.413 | -0.024 |
| **15°** | 0.524 | -0.113 | 0.294 | 0.589 | -0.430 | -0.284 | 0.270 | 0.096 | -0.101 |
| **30°** | -0.781 | 1.635 | 0.171 | -0.709 | -0.315 | 0.332 | 0.152 | -0.070 | -0.057 |
| **45°** | -0.030 | 0.646 | 0.019 | 0.015 | -0.167 | -0.009 | -1e-5 | 0.004 | 0.010 |
| **60°** | 0.090 | 0.473 | 0.173 | 0.173 | -0.432 | -0.128 | 0.132 | 0.044 | -0.018 |
| **75°** | -0.892 | 1.755 | 0.436 | -0.730 | -0.864 | 0.220 | 0.391 | -0.012 | -0.104 |

**Table 2.** Dominant $f_{lm}$ parameters evaluated for the best fit of the angular distribution presented in Fig. 12, over an energy range of 1 – 20 keV, a porosity range 0.49 – 0.87, and an incident angle range of 0° – 75°.

5. Conclusions

We have developed a sputtering model for loose powders structures that can also be applied to rough surface, pressed pellet, and fuzzy targets. The powder was described in MD using LAMMPS GRANULAR to account for physical grain-to-grain contacts. Additionally, this model combines a ray-tracing model with a Monte Carlo approach to describe the ion-target interactions, with the latter describing the sputtering yield direction as a function of the local incidence angle. We have applied this model to Kr ions impacting a Cu target for both a loose powder and a flat slab.

Our simulations show that ion sputtering from a loose powder is markedly different from sputtering from flat slabs from non-porous rough surfaces or pressed pellets and from foam-like fuzzy targets. We attribute these differences to the interconnected voids in a loose powder that are not present in these other samples. In specific, we find that:

- The escaped sputtering yield from a powder for $\alpha > 0°$ is predominantly in the backward direction (relative to the incident ion) for all impact energies.

- The percentage of the yield that effectively leaves the powder remains nearly constant with ion energy, but varies with the incidence angle and with the porosity. The peak in the angular distribution is typically half or less than that of a flat slab.

- The ejecta distribution does not show the evolution from primary knock-on sputtering to secondary knock-on sputtering that is seen in sputtering from non-porous targets for $\alpha > 0°$.

- Sputtering from a powder cannot be approximated by the sputtering of a flat slab nor non-porous rough surfaces and pressed pellets nor foam-like fuzzy targets.

Using our results, we have derived a potentially universal scaling relationship between the total escaping sputtering yield from a granular structure and that from a flat surface. This relationship depends only on the porosity $p$, the incident angle $\alpha$, the mean local incidence angle of the powder $\alpha'_M$, and the sputtering yield for an equivalent composition flat slab at this angle. Clearly additional theoretical and experimental work are needed to validate the proposed relationship. However, partial validation come from the fact that a similar relationship can be used to reproduce satellite measurements of H ENA reflection on the Moon, showing a potential to be applied to a broad range of ionic species and surface compositions. Additionally, we used our result to derive a fit to the angular distribution of the escaping yield from a powder. We show that goodness of the fit is negligibly affected by beam energies of 1 – 20 keV and sample porosity $p \geq 0.49$, enabling use of the fit over a wide range of porosity and energy. Combining the scaling relationship derived for the total sputtering yield with this angular distribution fit offers the potential of greatly simplifying modeling studies of the sputtering of powders, such as for planetary exospheres. As a final point, we note that the model developed here provides a higher-fidelity approach for the treatment of the target geometry as compared to the voxel approximation used in other models. In future work, we will investigate the effects of the grain size distribution, non-spherical grain shapes, surface roughness of the individual grains, and heterogeneous compositions of the grains.

Data Availability

The code and data used in this study are available at Verkercke et al. (2025). "Loose Powder Sputtering Simulation (LooPSS)", DOI [10.5281/zenodo.15129281](10.5281/zenodo.15129281).

Acknowledgments

D.W.S. thanks F. Aumayr, D. Domingue, and M. Sarantos for stimulating conversations. S.V. and F.L. acknowledge the support of ANR, France of the TEMPETE project (grant No. ANR-17-CE31-0016) and the support of CNES, France for the BepiColombo mission. This research was supported by the International Space Science Institute (ISSI) in Bern, through ISSI International Team project #616 "Multi-scale Understanding of Surface-Exosphere Connections (MUSEC)." The authors also acknowledge the support of the IPSL CICLAD data center for providing access to their computing resources and data. C.B and D.W.S were supported, in part, by the NASA Solar System Workings program under Grant No. 80NSCC22K0099.

Declaration of Competing Interests

The authors declare that they have no known competing financial interests or personal relationships that could have appeared to influence the work reported in this paper.

Author Contributions

**Sébastien Verkercke**: Conceptualization (equal), Data curation (lead); Formal analysis (equal); Investigation (equal); Methodology (equal); Software (lead); Validation (equal); Visualization (equal); Writing – original draft (lead); Writing – review & editing (equal). **Deborah Berhanu:** Formal analysis (supporting); Investigation (equal); Methodology (supporting); Writing – review& editing (supporting). **Caixia Bu**: Conceptualization (supporting), Data curation (supporting); Formal analysis (equal); Investigation (equal); Methodology (supporting); Software (supporting); Writing-original draft (supporting); Writing – review & editing (supporting). **Benjamin A. Clouter-Gergen:** Data curation (supporting); Formal analysis (supporting); Investigation (equal); Methodology (supporting); Software (supporting); Validation (supporting); Visualization (supporting); Writing – review& editing (supporting). **François Leblanc:** Funding acquisition (equal); Investigation (supporting); Methodology (equal); Resources (equal); Software (supporting); Supervision (supporting); Writing – review & editing (supporting). **Jesse R. Lewis:** Data curation (supporting); Formal analysis (supporting); Investigation (equal); Methodology (supporting); Software (supporting); Validation (equal); Visualization (supporting); Writing – review& editing (supporting). **Liam S. Morrisse**y: Conceptualization (supporting), Data curation (supporting); Formal analysis (equal); Funding acquisition (equal); Investigation (equal); Methodology (equal); Resources (equal); Software (supporting); Supervision (supporting); Validation (equal); Visualization (equal); Writing-original draft (supporting); Writing – review & editing (supporting). **Daniel. W. Savin:** Conceptualization (equal); Data curation (supporting); Formal analysis (equal); Funding acquisition (supporting); Investigation (equal); Methodology (equal); Software (supporting); Supervision (lead); Validation (equal); Visualization (equal); Writing-original draft (supporting); Writing – review & editing (equal).